\documentclass[12pt,preprint]{aastex}

\begin{document}

\title{Infrared Extinction Toward Nearby Star-Forming Regions}

\author{Flaherty, K.M.\altaffilmark{1, 2}, Pipher, J.L.\altaffilmark{2}, Megeath, S.T.\altaffilmark{3}, Winston, E.M.\altaffilmark{4}, Gutermuth, R.A.\altaffilmark{2, 4}, Muzerolle, J.\altaffilmark{1}, Allen, L.E.\altaffilmark{4}, Fazio, G.G. \altaffilmark{4} }

\email{kflaherty@as.arizona.edu}
\altaffiltext{1}{Steward Observatory, University of Arizona, Tucson, AZ 85721.  This work was completed when the first author was at the University of Rochester.}
\altaffiltext{2}{Department of Astronomy, University of Rochester, Rochester, NY 14627}
\altaffiltext{3}{Ritter Observatory, University of Toldeo, Toldeo, OH 43606}
\altaffiltext{4}{Harvard-Smithsonian Center for Astrophysics, 60 Garden Street, Cambridge, MA 02138}

\begin{abstract}
We present an independent estimate of the interstellar extinction law
for the {\it Spitzer} IRAC bands as well as a first attempt at extending
the law to the 24\micron\ MIPS band. The source data for these
measurements are observations of five nearby star-forming regions: the
Orion A cloud, NGC 2068/71, NGC 2024/23, Serpens and Ophiuchus. Color
excess ratios $E_{H-K_s}/E_{K_s-[\lambda]}$ were measured for stars
without infrared excess dust emission from circumstellar disks/envelopes.
 For four of these five regions, the extinction laws are
similar at all wavelengths and differ systematically from a previous
determination of the extinction law, which was dominated by the diffuse
ISM, derived for the IRAC bands. This difference could be due to the
difference in the dust properties of the dense molecular clouds observed
here and those of the diffuse ISM. The extinction law at longer
wavelengths toward the Ophiuchus region lies between that to the other
four regions studied here and that for the ISM. In addition, we extended
our extinction law determination to $24\micron$ for Serpens and NGC
$2068/71$ using {\it Spitzer} MIPS data.  We compare these results
against several ISO extinction law determinations, although in each case
there are assumptions which make absolute comparison uncertain. However,
our work confirms a relatively flatter extinction curve from 4 - 8$\mu$m
than the previously assumed standard, as noted by all of these recent
studies. The extinction law at $24\micron$ is consistent with previous
measurements and models, although there are relatively large
uncertainties.

\end{abstract}


\section{Introduction}

With the launch of the {\it Spitzer} Space Telescope, identification
of embedded young stellar objects (YSOs) from their infrared excess emission at mid-infrared wavelengths ($3-8\micron$ using IRAC, the Infrared Array Camera, and at 24$\mu$m, using MIPS, the Multiband Imaging Photometer) has been  facilitated by {\it Spitzer's} unprecedented sensitivity. These wavelengths are sensitive to the emission from disk and envelope material surrounding YSOs, making it easier to differentiate young stellar objects from foreground or background stars where the emission is primarily photospheric. \citet{all04} and \citet{mea04} develop a classification scheme to differentiate young stellar objects based on their IRAC colors, and \citet{muz04} present a scheme utilizing 24$\mu$m data in combination with IRAC data. \citet{gu05} has recently presented a robust classification scheme for YSOs which utilizes de-reddened mid-IR fluxes of the cluster members in combination with ground-based photometry.  These classification  schemes are briefly compared in the recent review article by \citet{allen07}. \citet{lada06} also employ de-reddened IRAC observations to identify YSOs in IC348, using an alternate classification scheme. Since these schemes involve de-reddening, it is critical to provide an accurate extinction law at IRAC and MIPS wavelengths that is appropriate for young clusters embedded in nascent molecular cloud material. 

Examining the extinction to background stars beyond these embedded young clusters probes the column density of dust in the molecular cloud. The availability of 2MASS has allowed for construction of large scale extincton maps in the near-infrared which trace a substantial range in column densities \citep{lada07,lom06}. With an appropriate extinction law, this technique could be extended to longer wavelengths where the extinction is lower and higher column densities can be probed.

Throughout much of the IRAC range of interest (0.7 - 5$\mu$m), \citet{dr89} fit data existing at that time to a power-law  $A_\lambda \varpropto \lambda^{-\beta}$, with $\beta$ = 1.75, and extrapolated the power law to meet the  beginning of a silicate contribution at $\sim$7 $\mu$m in his Fig. 1; other early papers have found $\beta$ ranging from 1.61 \citep{rl85} to 1.80 \citep{mw90,whi93}. Dust models \citep{wei01} employing size distributions of various grain components with known optical properties have been constructed to match observed extinction laws.  Measurements of the extinction law from $\sim0.9\micron$ to
$\sim2\micron$ have been reported by many investigators using a
variety of techniques (from both ground-based and space observations)
along a number of lines of sight, and these were thought to be
``universal", i.e. generally in agreement with each other
\citep{dr89,car89}, although recently some authors have questioned this
universality \citep{gk01,rgk02}. Longward of $2\micron$ the extinction is less certain  (see the range in $\beta$ above), and the most robust
estimates are generally obtained with instruments on spaceborne
telescopes. Pre-$Spitzer$ studies have relied primarily on
measurements of hydrogen recombination or hydrogen rotational and
ro-vibrational lines \citep{lut96,lut99,ros00,ber99} to estimate the
extinction law for $\lambda>2.5\micron$.  For example, \citet{lut99}
used hydrogen recombination line emission toward the galactic center
to probe the extinction from $2.5\micron < \lambda < 20\micron$ and
found that the extinction law $A_\lambda$/$A_V$ is flatter from
$4-8\micron$ than the power laws noted above, and approximately
$4\times$ smaller than the value of the standard curve at
7$\mu$m. \citet{lut99} notes that the lack of a deep minimum near
7$\mu$m is in conflict with predictions assuming the standard mixture
of silicate and graphite grains \citep{dr89}. 
\citet{ber99} and \citet{ros00} observed $H_2$
rotational and ro-vibrational lines toward OMC-1 and assumed a power
law ($\lambda^{-1.7}$ extending from 2.4 to $6\micron$), and fit two
silicate features to the extinction law beyond $8\micron$.  The latter
authors also determined extinctions from hydrogen recombination lines,
and obtained smaller values of the relative extinction by this method
than with $H_2$ lines. In that case $A_{\lambda}$/$A_V$ was also
relatively flatter from 3$\mu$m - 7$\mu$m, and from 1.5 - 2 times
smaller than the comparison curve (in this case the IR extinction law
from \citet{co93} with modifications to the 10$\mu$m silicate feature
strength: the Cohen law is very similar otherwise to the one adopted
by \citet{dr89}).  A study using ISOGAL, DENIS and 2MASS data of
intermediate AGB and RGB stars in the inner galactic plane by \citet{ji06}
determined values for $A_7$/$A_{K_s}$ and $A_{15}/A_{K_s}$ that are more
comparable to those of \citet{lut99} than to earlier `standard'
interstellar extinction curves in the solar neighborhood \citep{dr89}.

Using observations of stars they identified as reddened clump giant stars in the galactic plane
obtained as part of the {\it Spitzer} GLIMPSE survey, \citet{ind05}
directly measured the extinction law for the {\it Spitzer} IRAC bands
at 3.6, 4.5, 5.8, and 8.0$\mu$m. They observed the massive
star-forming region RCW 49 at a distance exceeding 2 kpc, as well as two off-field regions very close to the
galactic plane, and supplemented the {\it Spitzer} data with 2MASS
observations of the same fields.  For the off-field regions, their method determines the diffuse 
interstellar medium extinction, assuming that the red clump giants are uniformly distributed along the line of sight. Reddened red clump giant stars were
selected from their position in the $J$ vs. $J-K_s$ color-magnitude
diagram, and a reddening band was identified which was then used to
determine $A_J/A_{K_s}$ assuming a narrow distribution of intrinsic colors and luminosities of the clump giants. $A_{\lambda}/A_{K_s}$ was then determined from $[\lambda]-K_s$ vs. $J-K_s$ color-color diagrams for these objects. The extinction curve
for all three of these regions was nearly identical and also appeared
to have a shallower slope for $\lambda>4\micron$ than the standard
law.  R. Indebetouw kindly provided the data for these three regions so that we could process them using the same methods that we used to process our data, hence making possible a direct comparison of the extinction laws toward the young clusters studied here, and that along the lines-of-sight probed by \citet{ind05}.

Here we determine the extinction law from $3-8\micron$ toward five young, nearby ($<$ 1 kpc) cluster regions, including the Orion A cloud, NGC 2068/71, NGC 2024/23, Ophiuchus and Serpens.  We further determine the extinction at $24\micron$ for two of the regions, NGC 2068/71 and Serpens.  In all cases we measure extinctions in relatively broad wavebands as determined by fixed, cold filters.  By examining those stars without disk/envelope infrared excess emission in the cloud direction, we evaluate the selective extinction at mid-infrared wavelengths: the extinction derives primarily from their host molecular clouds, although extinction from the diffuse interstellar medium is also a factor.  We compare our determination against prior determinations in the literature including the ``standard extinction law" such as that compiled by \citet{dr89}. It is important when comparing these determinations to keep in mind that spectroscopic measurements may differ from our broad-band measurements at these wavelengths.

Our separate extinction law study at the wavelengths of the IRAC bands
is motivated by a discrepancy between slopes of the bands of reddened
sources in the $H-K_s$ vs. $K_s-[\lambda]$ color-color diagrams for our
clusters, and slopes found using the GLIMPSE data. Reddened non-excess sources form a band on the $H-K_s$
vs. $K_s-[\lambda]$ color-color diagrams extending from the
intrinsic colors of the stars parallel to the reddening vector whose
{\it slope} is the color-excess ratio $E_{H-K_s}/E_{K_s-[\lambda]}$ $\equiv$ $(A_H- A_{K_s})/(A_{K_s}-A_{\lambda})$. These color excess ratios, derived in  Section~\ref{obs}
and given in Table 2 for our data, and for the GLIMPSE data provided by
Indebetouw, illustrate at a glance the discrepancy between selective extinction slopes toward
our sources and theirs. \citet{hu07} also found a slope discrepancy between the selective extinction law toward isolated dense molecular cores observed in the {\it Spitzer} Cores to Disks (c2d) Legacy program \citep{ev03}, and toward those regions examined by \citet{ind05}.

For these young clusters, and for the regions examined by \citet{ind05}, we utilize in our analysis those stars without a mid-infrared excess in the IRAC color-color diagrams \citep{all04}, namely class III cluster objects, and field stars. Background dwarfs and red giants, including the red clump giants, are the primary probes of the extinction that originate in the clusters' nascent molecular clouds. Using complementary 2MASS ground-based $H$ and $K_s$ data as well as $Spitzer$ data on these objects, we measure the color excess ratios $E_{H-K_s}/E_{K_s-[\lambda]}$ from $H-K_s$ vs. $K_s-[\lambda]$ color-color diagrams, and from these ultimately derive $A_{\lambda}/A_{K_s}$ for an assumed value of $A_H/A_{K_s}$.  The same method allows extension to 24$\mu$m for the Serpens and NGC $2068/71$ clusters, although these results are more uncertain because of the comparative paucity of detected sources. 

In Section~\ref{obs} the observations and data reduction will be discussed, and
Section~\ref{measure_ext} will describe source selection and the derivation of the
reddening law from these data. Section~\ref{results} will report the
results and discuss the differences between the extinction law derived
here for our data and for the data used in the study by \citet{ind05}, as well as the extinction law derived using different methods and extinction probes by \citet{ind05}, with further reference to other determinations \citep{lut99,ji06,ros00,hu07}.  We have not determined the total to selective extinction in this study, so shall concentrate on relative extinctions, assuming a value for $A_H/A_{K_s}$ in order to make comparisons with other work.  Because a non-universal extinction law may lead to variations in this assumed value, we consider how this assumption affects our results.

\section{Observations and Analysis}\label{obs}
Observations of several young star-forming clusters were obtained with
the IRAC camera \citep{faz04} aboard {\it Spitzer} as part of several surveys concentrating on star-formation within $1$ kpc of the Sun, including the ``Young Cluster Survey" (PID 6), the ``Orion Survey" (PID 43) from G. Fazio's guaranteed observation time, as well as the ``Cores to Disks Legacy Program" (PID 174 and 177). The IRAC camera takes
simultaneous images in four channels whose isophotal wavelengths are
$3.550$, $4.439$, $5.731$ and $7.872\micron$ respectively
\citep{faz04}. Some of the regions observed cover several square degrees so
that multiple AORs with the IRAC camera were required. NGC
$2024$ and NGC $2023$ were imaged as part of the same region, while NGC
$2068$ and NGC $2071$ were also observed together. Observations of the
Orion A cloud included coverage of OMC 2/3, ONC and L1641. High
dynamic range mode \citep{faz04} was used, resulting in a frame with a
$0.4$ second exposure time and a frame with a $10.4$ second exposure
time at each position. The maps for each of NGC 2024/23, NGC 2068/71 and Orion A were repeated twice. The maps of Serpens and Ophiuchus were repeated four times, with small offsets. The additional dithers provide both additional sensitivity to faint sources and greater ability to remove cosmic rays from the final mosaics. We have not corrected for the position dependent gain variation for point sources\footnote{See \url{http://ssc.spitzer.caltech.edu/irac/locationcolor/}}, which would result in an additional 5\% uncertainty. 

For the Orion fields individual frames were mosaiced using a custom
IDL program and from these mosaics, source selection and point source
photometry were completed using Gutermuth's PhotVis v$1.08$
\citep{gut04}. The photometry used an aperture of $2.4\arcsec$ and a
sky annulus with inner and outer radii of $2.4\arcsec$ and
$7.2\arcsec$. Details on the data reduction and source selection in
Serpens and Ophiuchus are described in \citet{wi07} and
\citet{all07}. \citet{Reach05} discusses the calibration of the IRAC
instrument. The calibration of the {\it Spitzer} telescope was refined during the course of this work, and the resulting zero points are listed in Table 1. However, changes in the zero point do not change the results in this paper and are taken into account in our fitting technique. From the 2MASS point source database we obtained
corresponding $J, H$ and $K_s$ band magnitudes for each star. For the
NGC $2068/71$, NGC $2024/23$ and Serpens regions, we required
magnitude uncertainty $\sigma<0.1$ in all seven bands for sources used
to calculate the extinction law. For the Orion A cloud and Ophiuchus a
detection in J band was not required, thereby increasing the number of
highly reddened sources and providing a better determination of the
extinction law. For the remaining 6 bands, a magnitude uncertainty of
$\sigma < 0.1$ was imposed. If we do not require a J band detection for NGC $2024/23$ , NGC $2068/71$ or Serpens, our derived values for the slope of the extinction vector are within $1\sigma$ of the values derived with the seven-band data.

Requiring sources to have this magnitude uncertainty 
reduced the spread of the sources on each color-color diagram, and
minimized the uncertainty in the fit (see Section~\ref{measure_ext}). Those sources for which only upper limits to 2MASS magnitudes were available were not included in our analysis.  Most of the sources had much smaller magnitude uncertainties. The median magnitude uncertainties for NGC 2024/23 sources 
are $0.03, 0.02, 0.02, 0.005, 0.005, 0.02, 0.05$ for $J, H, K_s$, and
channels 1 through 4 of IRAC respectively. Similar median magnitude uncertainties were obtained for the other regions studied here. Loosening the requirement on the maximum uncertainty will include more sources with lower fluxes. It is possible that these sources have higher extinction and could provide a more accurate measure of the reddening law.  However, increasing the uncertainty limit leads to the inclusion of more outliers that skew the slope and are difficult to remove with a simple color cut. A small number of fits did not have any outliers and for these fits including the sources with higher uncertainty does not change our results by more than the measured uncertainty in the color excess ratios.

MIPS 24 $\mu$m images were available for several of the regions
studied. Not all regions were suitable - there were insufficient non-excess sources in some regions to reliably derive an extinction law.  We report on NGC 2068/71 and Serpens observations with the Multiband Imaging Photometer for
{\it Spitzer} \citep[MIPS,][]{ri04}. We summarize the MIPS observations, with more detailed descriptions of the data reduction and photometric measurements for Serpens and NGC 2068/71 in \citet{wi07} and \citet{muz07} respectively. The measurements were obtained in scan mode with half-array cross-section offsets. All three MIPS bands are observed simultaneously in this mode; the total effective exposure time per point is about 80 seconds at 24 microns. Photometry was determined using a PSF fitting program in IDL.

\section{Background Source Selection and Determination of Extinction Law}\label{measure_ext} 

% We now describe how we selected suitable background extinction probes.
 Background dwarfs and giants (including red clump giants and RGB stars) as well as embedded pre-main sequence stars can be used to measure the extinction due to the molecular cloud dust
associated with the star-forming region as well as diffuse
line-of-sight extinction. As mentioned above, the sources used to measure the extinction were chosen from 
those without infrared excesses from disk or envelope emission in the
$[3.6]-[4.5]$ vs. $[5.8]-[8.0]$ color-color diagram toward each cluster
region. Eliminating infrared excess sources from
consideration removes any ambiguities between stars with red colors
caused by extinction vs. stars with circumstellar dust or envelope
emission, including both pre-main sequence stars and AGB stars. A non-excess source was defined as having colors
$[3.6]-[4.5] < 0.6$ and $[5.8]-[8.0] < 0.2$. This region of
consideration was chosen so that it did not overlap with the expected
position of pre-main sequence stars with disks according to \citet{all04} and included
the expected spread in colors due to interstellar dust reddening,
which is larger at the shorter wavelengths. This definition of
non-excess sources is consistent with the theoretical models of
\citet{whi03}, and the observations of \citet{mea04} and
\citet{gut04}. An upper limit on the $[3.6]-[4.5]$ color is used to
eliminate unresolved outflow sources misidentified as stars. Outflows
have emission features in the 4.5 $\mu$m band, leading to
$[3.6]-[4.5]$ colors which could be confused with those from reddened
non-excess stars. 

We extend our selection criteria to attempt to further remove foreground stars, and stars whose colors place them outside the reddening band. We remove foreground giants by excluding any star brighter than a K2III giant with $M_J=-0.9$ \citep{lop02} at the distance of each cluster. We also exclude stars with $H-K_s<0.2$, which are mostly foreground dwarfs (see discussion below). Stars with $K_s-[3.6]<0$ do not fit into the reddening band and are removed as well as stars that appear to have an excess in the $H-K_s$ vs $K_s-[8.0]$ diagram (e.g. $H-K_s< 1.04(K_s-[8.0])-0.2$ for Serpens). We have not attempted to use only clump red giants, as was done by \citet{ind05}, because of the difficulty in distinguishing these stars from highly reddened dwarfs and pre-main sequence stars without disks. Our reddest stars will be dominated by red giants and embedded pre-main sequence stars because of the large intrinsic luminosity needed to detect these stars at such high extinction although at lower extinction we may have a mix of reddened dwarfs as well as giants and pre-main sequence stars. For Orion A, a number of galaxies contaminated the field and their removal from consideration is discussed in section~\ref{oriona}.

\begin{figure}
\epsscale{1.}
\plotone{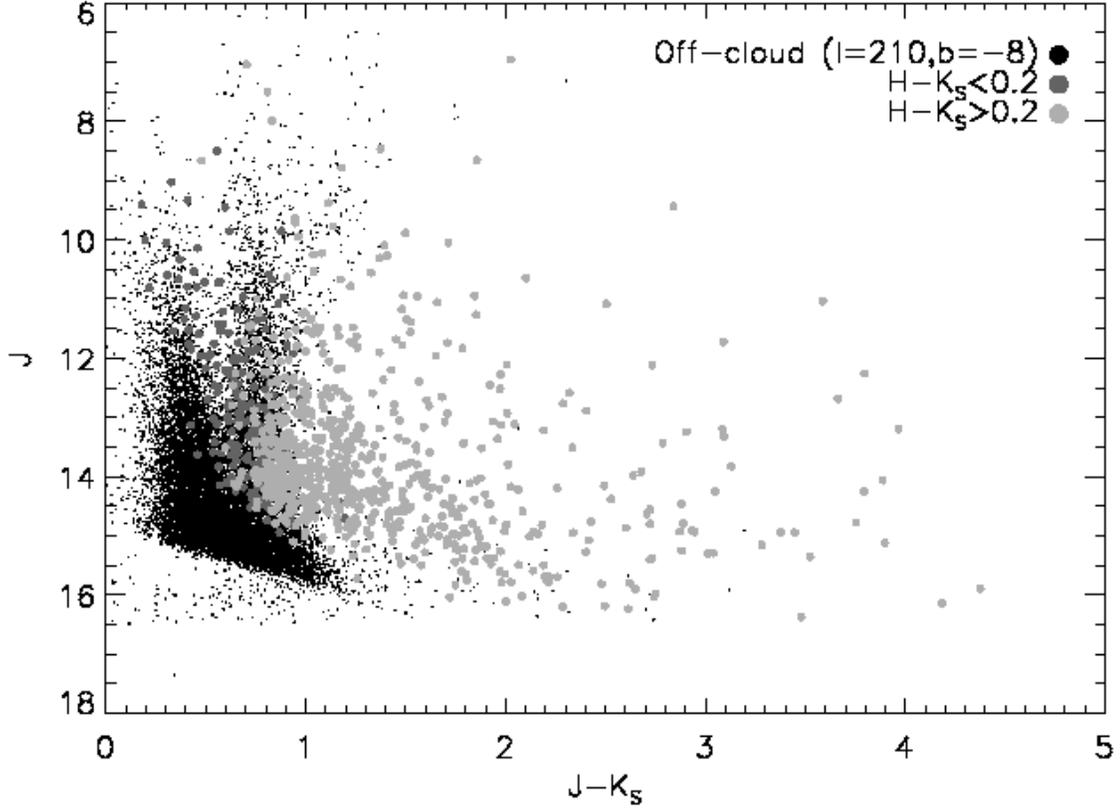}
\caption{$J$ vs. $J-K_s$ diagram for the NGC 2068/71 cluster (plotted in  blue), and an equivalent area off-field region (black).  Red symbols represent cluster stars excluded by the $H-K_s$ cutoff. For the off-cloud (black symbols) line of sight, dwarfs occupy the peak at $J-K_s$=0.2, while red clump sources are distributed around $J-K_s$=0.75.  On the other hand, toward the clusters (blue) red clump extinction probes are found peaked around $J-K_s$=1.05 at $J$ = 14.5 toward the cluster. The solid line is a reddening vector for $A_V=5$ extending from $J=14.5$,$J-K_s=0.75$ \citep{rl85}.\label{JvJ-K_2068_off}}
\end{figure}

\citet{lop02} point out that stars lying to bluer colors than the red clump giant distribution in a $J$ or $K$ vs. $J-K$ color-magnitude diagram for a given line of sight are primarily dwarfs, while stars lying to redder colors are predominantly M giants and AGB stars. The red clump giant distribution is almost vertical around a median $J-K$ = 0.75 $\pm$ 0.2 for off-cloud lines of sight at galactic latitudes with absolute value $\ge$ 6$^{\rm o}$. For fields probing greater extinction from either molecular clouds or along lines of sight closer to the galactic plane, the red clump giant distribution shifts to redder $J-K$, fainter magnitudes and the distribution is broader. All of our clusters have galactic latitudes exceeding 6$^{\rm o}$, with the exception of Serpens, at $b$=5.4$^{\rm o}$. To illustrate the above points for one of our clusters, we show a 
$J$ vs. $J-K_s$ diagram for NGC 2068/71 (a region of high extinction) as well as an off-cloud region overplotted (Fig.~\ref{JvJ-K_2068_off}).  The off-cloud field is centered at l=210$^{\rm o}$, b=-8$^{\rm o}$, a region that was chosen because it did not show any significant CO emission indicative of a dense molecular cloud \citep{wi05}. In that diagram, the sources coincident with the cluster utilized in measuring the color-excess ratios are plotted with blue symbols,  and red symbols illustrate cluster sources rejected by the $H-K_s$ cutoff.  Off-cluster sources are plotted with black symbols. In the black double-peaked distribution, dwarfs at various distances along the line of sight occupy the bluer peak, and the red clump sources the redder, with median color $(J-K)_{\rm o}$ = 0.75, as noted above.  The effect of the molecular cloud's reddening can be seen by noting that most red clump extinction probes to NGC 2068/71 form a clump centered at $J-K_s$=1.05 at $J$=14.5. The off-cloud giants have nearly constant $J-K_s$ color, indicating little extinction from the diffuse ISM. For this reason we expect the contribution of the diffuse ISM to our measured extinction law for the Orion fields to be small, although this may not be equally valid for Serpens and Ophiuchus. Many sources redder than the red clump giants are sampled, and in particular, for the cluster region (red and blue points) in Figure~\ref{JvJ-K_2068_off} there is no distinct red clump giant locus. In the cluster region, the highly variable extinction makes it impossible to isolate red clump giants, reddened dwarfs and embedded pre-main sequence stars. We are likely dominated by red clump giants, and perhaps pre-main sequence stars, in the densest parts of the cluster, because they are bright enough to be observed at large reddening. However, we do not attempt to use only red clump giants.

We now move to discussion of reddening law determination. A straight line was fit to
background sources in the `reddening' band on an $H-K_s$
vs. $K_s-[\lambda]$ color-color diagram. After all of the restrictions described above were imposed,  there were 415, 670, 1830 and 1225 background sources for NGC $2023/24$, NGC $2068/71$, Orion A and Serpens respectively. Ophiuchus had 120 background sources and the implications of this small sample size are discussed further in section~\ref{results}. The slope of the line fit to the `reddening' band gives the color-excess ratio
$E_{H-K_s}/E_{K_s-[\lambda]}$ which is then used to determine the
relative extinction law $A_{\lambda}/A_{K_s}$, where we assume
$A_H/A_{K_s}$ = $1.55\pm0.08$ (the value derived by \citet{ind05}). The effect of a different choice of $A_H/A_{K_s}$ is discussed in section ~\ref{iso}.

The line fit to the reddening band data in each of the $H-K_s$
vs. $K_s-[\lambda]$ color-color diagrams minimized the chi-squared,
weighted by the uncertainty in each color.

\begin{equation}\label{chisq}
\chi^2(a,b)=\sum^{N}_{i=1}\frac{(y_i-a-bx_i)^2}{\sigma_{yi}^2+b^2\sigma_{xi}^2}
\end{equation}

\noindent where $a$ and $b$ are the y-intercept and slope, respectively,
of the best fit line and $\sigma_{xi}$ and $\sigma_{yi}$ are the
uncertainties in the x and y-values respectively. The algorithm {\it
amoeba} in IDL was used to minimize the chi-squared. Determining the
best fit-line to the reddening band data using equation~\ref{chisq} not only
accounts for the magnitude uncertainties, but also provides a method
for estimating the uncertainty in the slope, which leads directly to $A_{\lambda}/A_{K_s}$, the quantity of interest.

\citet{nr92} presented a detailed analysis of the adopted fitting technique as well as the characterization of its uncertainty, which we summarize here for the convenience of the reader. At the best-fit points the $\chi^2$ is a minimum, but if the values of
$a$ and $b$ are perturbed away from the best-fit values $\chi^2$
increases. A constant value of $\Delta\chi^2=\chi^2-\chi^2_{min}$
gives a confidence region within parameter space, which depending on
the size of $\Delta\chi^2$, could enclose a $68\%$, $90\%$,
etc. confidence interval. The uncertainty in $a$ and $b$
is then the projection of this confidence region onto the $a$ and $b$
axis. Finding $a,b$ where $\Delta\chi^2$ has a specific
value is determined analytically with a Taylor expansion when $\chi^2$
is linear in $b$, but must be determined numerically here because
$\chi^2$ is non-linear in $b$. This determination is performed by
varying $b$, minimizing $a$ at each step, until $\Delta\chi^2$ equals
the desired value. There are two values of $b$, one greater and one
less than the best-fit value, for which $\Delta\chi^2$ equals the
desired value. The distances of these two new values of $b$ from the
best fit slope are combined to determine the uncertainty in the
slope.

For the majority of our analysis $\Delta\chi^2$ was chosen to be 1,
which corresponds to a $68\%$ confidence level
\citep{nr92}. Increasing the confidence level to $99.99\%$ requires
$\Delta\chi^2=15.1$, which results in an increased uncertainty in the
slope. Both a $68\%$ and $99.99\%$
confidence interval were determined and are included in
Table~\ref{slopes}. The uncertainties are $\sim4$ times bigger for the
larger confidence interval than for the smaller confidence interval,
but lead to at the most a $6\%$ uncertainty in the fit.

The process that was used to derive the extinction law for the IRAC
bands can be applied to the $24\micron$ channel of MIPS. There were only sufficient numbers of non-excess sources with corresponding detections in 2MASS, IRAC and MIPS in the Serpens and NGC 2068/71 regions. The non-excess
sources were defined as having $[3.6]-[4.5]<0.6$, 
$[5.8]-[8.0]<0.2$ and $[8.0]-[24]<0.5$, with the last color cut used to remove sources that had an excess only at 24\micron. For both NGC $2068/71$ and Serpens fewer than 50 sources were
used for the final fit, making sure to exclude outliers and sources
with $H-K_s <0.2$, so we are working with quite small numbers of
sources. For this reason, we view these results with extreme caution.

\subsection{Orion A}\label{oriona}
While the process for deriving the extinction law from Serpens, Ophiuchus, NGC
2068/71 and NGC 2024/23 data was straightforward, the number and variety of
sources toward Orion A required more careful attention. The Orion A
cloud covers $4.4$ square degrees, a much larger area than the other four regions observed, increasing the number of sources used when
determining the reddening law but also increasing the number of outliers. These outliers are most obvious on the $H-K_s$
vs. $K_s-[8.0]$ color-color diagram on which there is a `branch' with
$H-K_s\simeq0.25$ which could skew the fit.

This `branch' is not as dominant in any of the other $H-K_s$
vs. $K_s-[\lambda]$ color-color-diagrams and has nearly constant
$H-K_s\simeq0.25$ color as well as $J-K_s\simeq0.9$. The objects in this branch are too blue in $H-K_s$ color to be infrared excess sources but they are too red in $K_s-[8.0]$ color to be main-sequence stars, red giants, or Class III
sources. The most likely explanation is that these sources are
galaxies. Their blue colors in the near infrared likely derive from
main-sequence stars dominating the light at short
wavelengths. Polycyclic Aromatic Hydrocarbon (PAH) emission from the
dust within the galaxies is the likely cause of the red $K_s-[8.0]$
color, since strong PAH feature emission falls within the $[8.0]$ band
of IRAC. Interpreting these sources as galaxies is consistent
with their fairly uniform spread in position throughout the Orion
A cloud. They are also not present in the areas with the
highest concentration of young stellar objects because these areas
have a large density of nebulosity which decreases the sensitivity for
detection of background objects, and a high column density of dust obscuring faint background sources. It is also possible that foreground sources contaminated by extended PAH emission are contaminants: however, this does not explain the lack of outliers in the regions of largest PAH emission near the young stars.  Regardless of the physical nature of these outliers, we remove them from our fitting procedure. A first attempt at removing these outliers was performed by changing the selection criteria for non-excess sources. This did not effectively remove the outliers without also removing a signifcant number of sources in the reddening band which are needed to accurately measure the extinction law. Removing these outliers, without removing a substantial number of sources in the reddening band, was completed by first fitting a line in the $H-K_s$ vs. $K_s-[8.0]$ color-color diagram to all of the
non-excess sources, including the outliers, and then using the slope of this fit to define a
boundary line ($H-K_s=1.11(K_s-[8.0])-0.2$), below which were assumed outliers and above which were
the sources used for a more accurate measure of the extinction. The
y-intercept of this line was chosen so that a majority of the outliers
was removed without removing a significant portion of the reddening
band. The initial fit in the $H-K_s$ vs. $K_s-[8.0]$ color-color diagram had a slope of $1.110$ and the final fit had a
slope of $1.116$ with $\chi^2/N$ decreasing from $1.99$ for the initial
fit to $1.48$ for the final fit. While this is a small correction, the
final fit is a more accurate measurement of the extinction law at
$8\micron$.

\section{Results}\label{results}
\subsection{The Extinction Law from 3-8$\mu$m}\label{extlaw}

\subsubsection{IRAC-Derived Color Excess Ratios}
\begin{figure}
\epsscale{1.}
\plotone{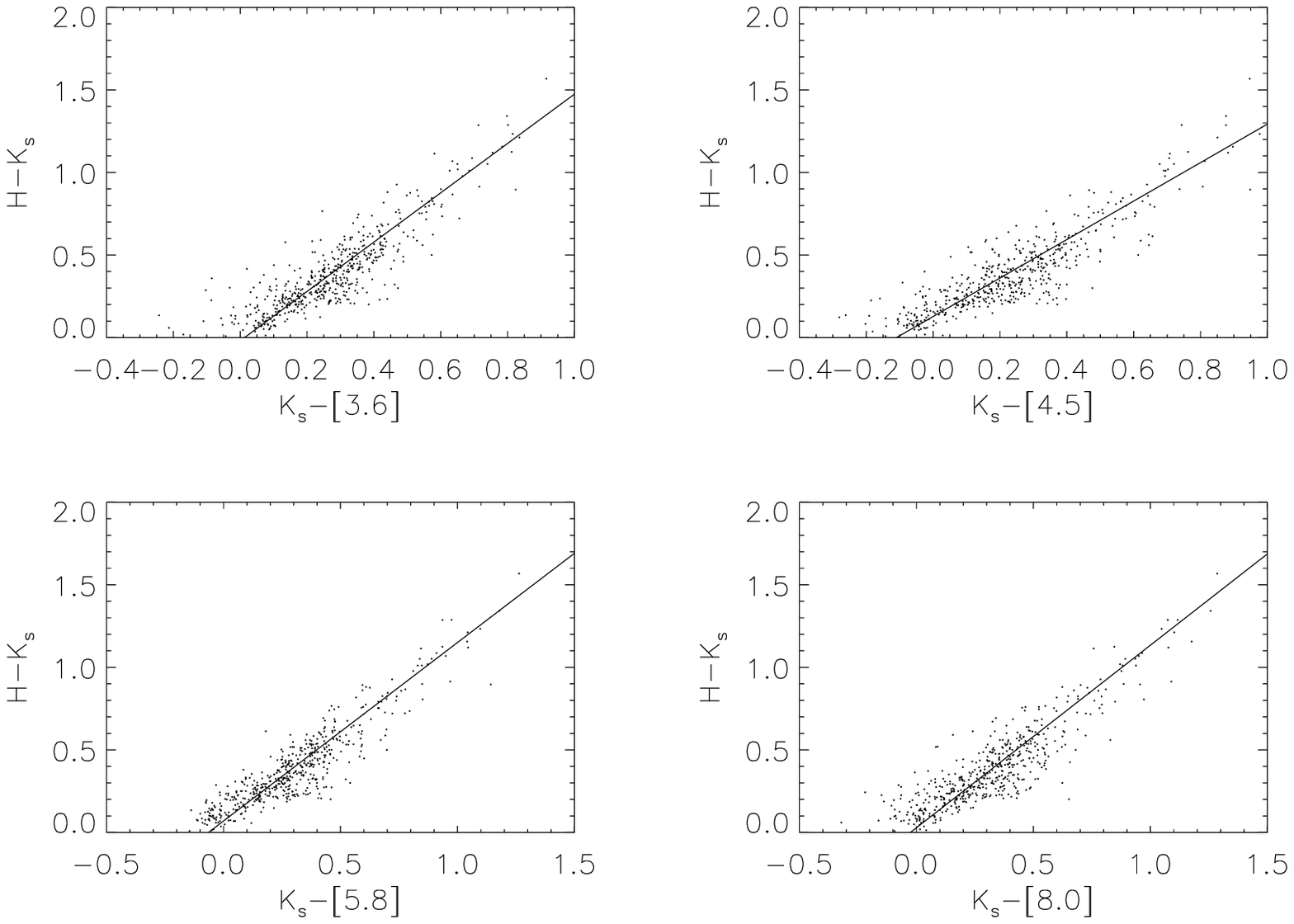}
\caption{2MASS and IRAC color-color diagrams for NGC $2024/23$ non-excess sources. Line is the best fit to the
data. Sources with $H-K_s<0.2$ or $K_s-[3.6]<0$ were not included in the
fit. Slopes are listed in Table~\ref{slopes}\label{ngc2024_red}.}
\end{figure}

\begin{figure}
\epsscale{1.}
\plotone{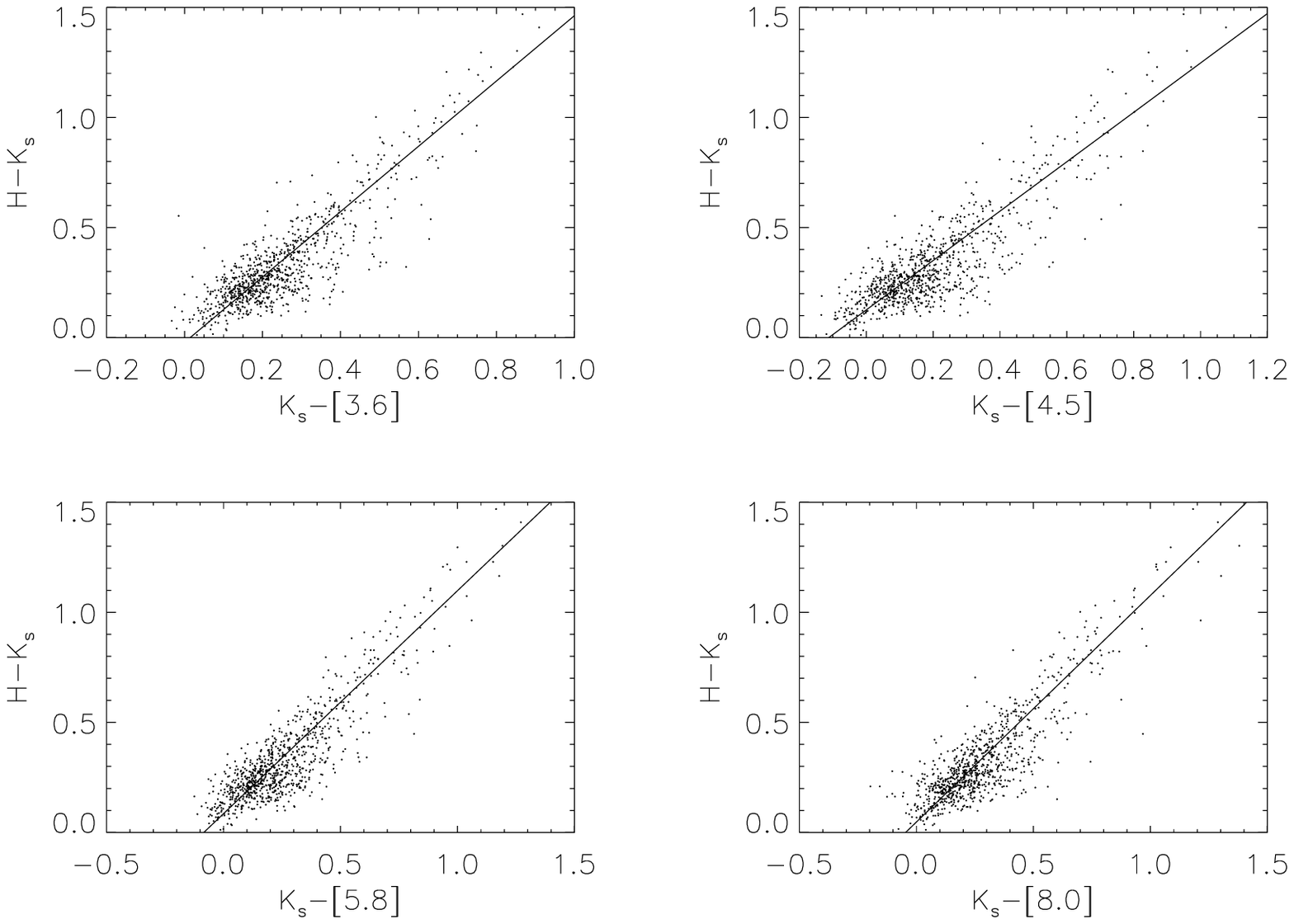}
\caption{2MASS and IRAC color-color diagrams for NGC $2068/71$ non-excess sources. Line is the best fit to the
data. Sources with $H-K_s<0.2$ or $K_s-[3.6]<0$ were not included in the
fit. Slopes are listed in Table~\ref{slopes}\label{ngc2068_red}.}
\end{figure}

\begin{figure}
\epsscale{1.}
\plotone{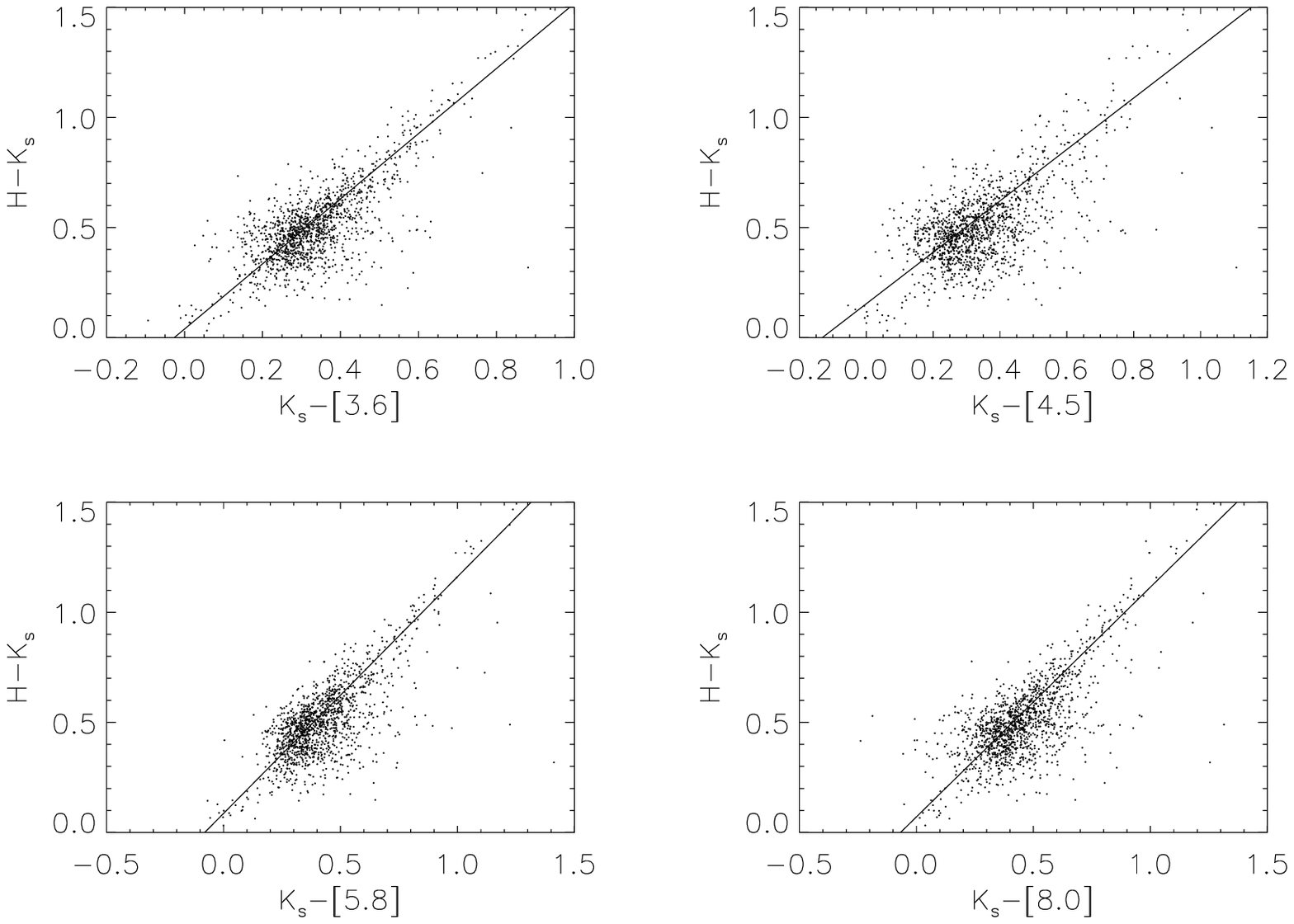}
\caption{2MASS and IRAC color-color diagrams for Serpens non-excess
sources.  Line is the best fit to the data. Sources with $H-K_s<0.2$ or
$K_s-[3.6]<0$ were not included in the fit. Slopes are listed in
Table~\ref{slopes}\label{serp_red}.}
\end{figure}

\begin{figure}
\epsscale{1.}
\plotone{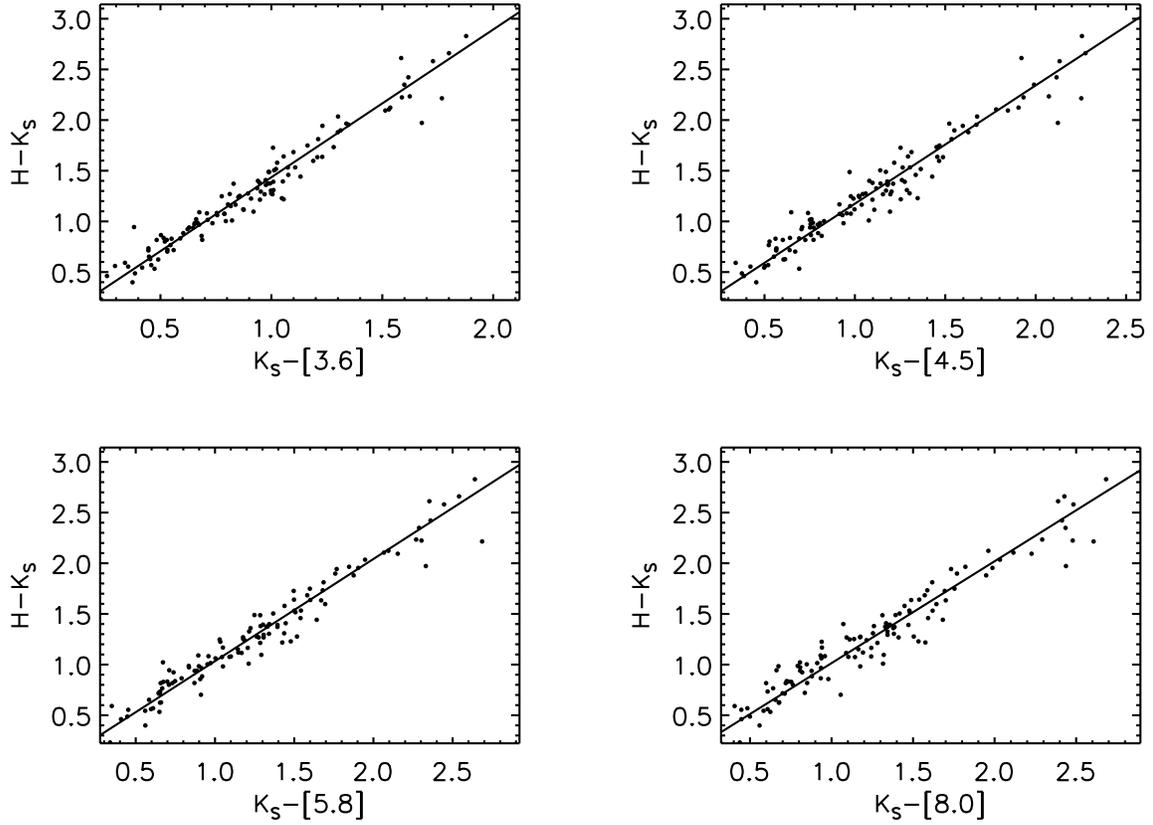}
\caption{2MASS and IRAC color-color diagrams for Ophiuchus non-excess sources. Line is the best fit to the data. Slopes are listed in Table~\ref{slopes}.\label{oph_red}}
\end{figure}

\begin{figure}
\epsscale{1.}
\plotone{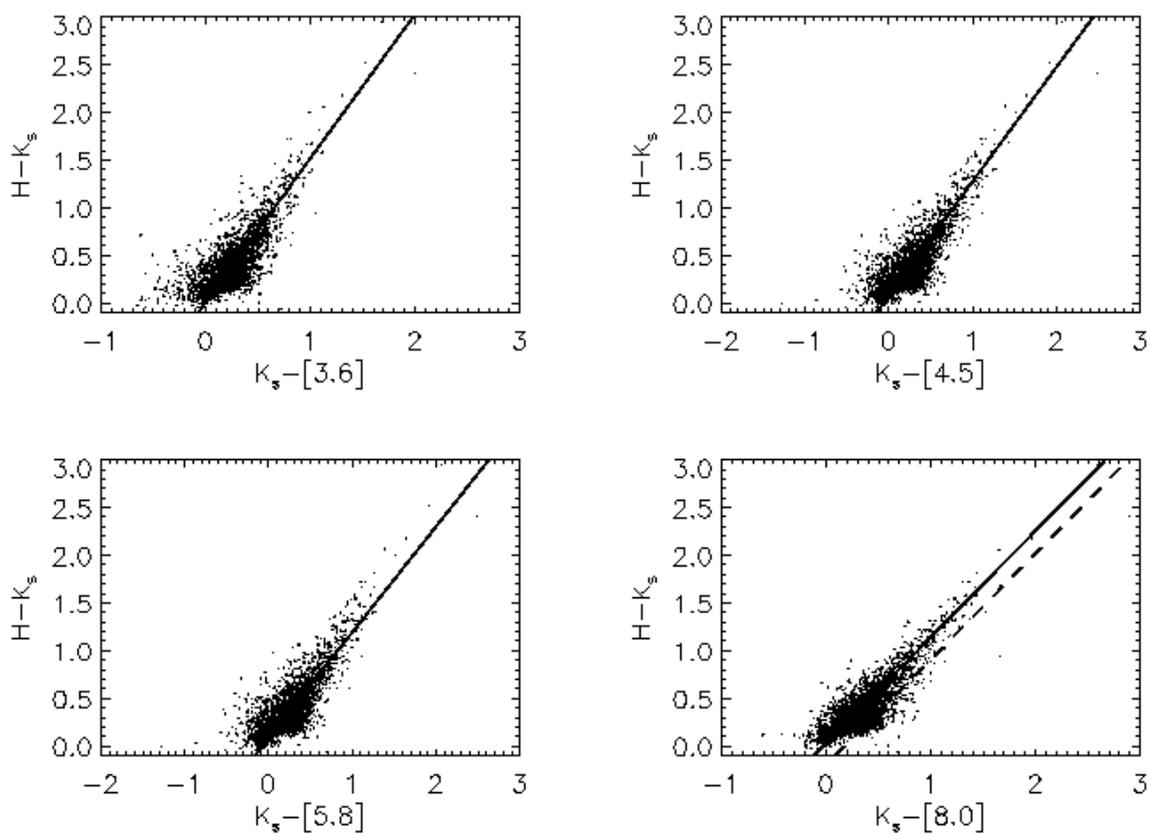}
\caption{2MASS and IRAC color-color diagrams for Orion A non-excess sources.  Line is the best fit to
the data. Sources with $H-K_s<0.2$ or $K_s-[3.6]<0$ were not included in
the fit. Slopes are listed in Table~\ref{slopes}. The dashed line is the boundary used to eliminate the branch near $H-K_s=0.25$.\label{oriona_red}.}
\end{figure}

The slopes of the linear fits to the $H-K_s$ vs. $K_s-[\lambda]$ color-color
diagrams for our five clusters are listed in Table~\ref{slopes} and
plotted in Figures~\ref{ngc2024_red}, ~\ref{ngc2068_red},
~\ref{serp_red}, ~\ref{oph_red}, and ~\ref{oriona_red}. It is apparent
that our derived slopes (color excess ratios) are reasonably
consistent from region to region, but in a few instances, are more
than 1-$\sigma$ different from each other at the longer
wavelengths. In particular, the selective color excess ratio toward
the Ophiuchus cloud is lower than toward the other four clouds at
8.0$\micron$. Since Ophiuchus has the smallest number of non-excess sources
(120 vs. $>$400 for the other four regions)we assess the possibility that the difference is due to the
small number of non-excess sources.  We randomly select 120 non-excess
sources from the Orion A field and measure the color-excess ratios in
the four IRAC bands for the subsample, repeating this process for 1000
subsamples. We can estimate how likely it is to measure the Ophiuchus
color-excess ratios based on the distribution of color-excess ratios
in each of the four bands for the subsamples. For the 5.8 and 8\micron\ bands the
average color-excess ratio for the subsamples is 1.115 and 1.122, with
a standard deviation of 0.06 for both bands. The Ophiuchus
color-excess ratios are approximately 2$\sigma$ from these mean
color-excess ratios at both 5.8 and 8\micron, only a marginally significant
deviation. The fraction of subsamples with color-excess ratios at or below the Ophiuchus values are 3\% and 1\% for the 5.8 and 8\micron\ bands respectively. Thus sample size is unlikely to be the cause of the lower selective color excess ratio at 8.0$\micron$ for Ophiuchus.

The color excess ratios for the $l=284^{\circ}, b=0.25^{\circ}$ 
off-cloud line-of-sight GLIMPSE data were derived using the same
source selection criteria and methods as used for the five
star-forming regions discussed here. In contrast to the general
agreement among the clusters, the slopes from the GLIMPSE data listed
in Table~\ref{slopes} are generally different from the cluster
slopes. For completeness, we also quote the color excess ratios from
\citet{ind05} for the $l=284^{\circ}$ off-cloud line of sight. Since
they directly measured $E_{[\lambda] - K_s}/E_{J-K_s}$, we form
$E_{H-K_s}/E_{K_s-[\lambda]}$ from those ratios in Table~\ref{slopes}.
Those values differ at 3.6 and 4.5 $\mu$m from values
determined by our analysis.

The question remains whether the larger sample size employed by \citet{ind05} in determining the extinction law is the cause of the discrepancy, as was suspected for Ophiuchus. Their method uses red clump giant stars selected by their $J-K_s$ colors and only requires a detection in the three bands ($J,K_s,[\lambda]$) that are used when measuring the color-excess ratios. This means that more sources are available for determining the color-excess ratios. To avoid this, and any other possible biases due to differing methodologies, we have reanalyzed the \citet{ind05} data using the same methods that were applied to our five clusters as noted above. The $l=284^{\circ}, b=0.25^{\circ}$ off-cloud line-of-sight was studied in detail because it had the largest number of background sources and its extinction law was representative of the other two lines of sight. There were 1590 background sources that met our criteria, comparable to the number of background sources used for four of our five clusters. \citet{ind05} find 3560 red clump giants along this same line of sight when requiring a detection in J and $K_s$ band. This includes stars that have an infrared excess due to circumstellar dust and stars that are not detected by IRAC. The actual number of sources used by \citet{ind05} when measuring the color-excess ratios was smaller ($<3000$) after requiring an IRAC detection and removing stars with an infrared excess. We examine the effect of differing sample sizes by considering the Orion A field and varying the limit on the magnitude uncertainty. When limiting the magnitude uncertainty to 0.1, there were 1825 background sources, while changing the limit from 0.05 to 0.25 results in sample sizes ranging from 1265 to 2360. The color-excess ratios for the extreme cases differ from the case where the magnitude uncertainty is less than 0.1, and from each other, by less than the measured uncertainty in the color-excess ratios.  As a result we do not expect differing sample sizes to provide a source of systematic error when comparing our results to those of \citet{ind05}.

\begin{figure}
\epsscale{1.}
\plotone{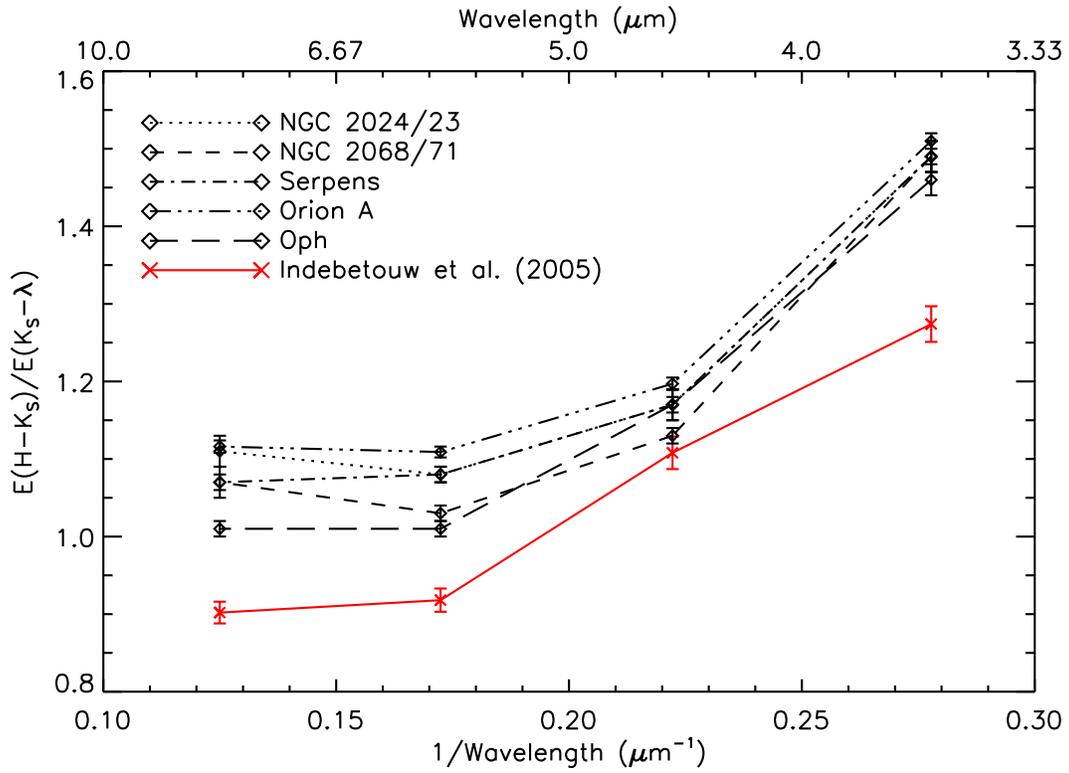}
\caption{Selective Color Excess Ratio versus inverse wavelength for
NGC $2024/23$, NGC $2068/71$, Serpens, Orion A, Ophiuchus and from the 
$l=284^{\circ}$ off-cloud region examined by \citet{ind05}, derived using the same method as for the five star-forming regions.\label{excess_wav}}
\end{figure}

Figure~\ref{excess_wav} gives the color excess ratio
$E_{H-K_s}/E_{K_s-[\lambda]}$ versus $1/\lambda$ for the five regions
observed here as well as for the GLIMPSE $l =284^{\circ}$ off-cloud
line-of-sight data analysed by us, showing the systematic separation
between the color excess ratios derived for the GLIMPSE line of sight,
and for the five young clusters. Our values are consistently 1-$\sigma$, or more, higher than those rederived from \citet{ind05} data with the exception of the $4.5\micron$ band.

The IRAC non-excess stars used when measuring the color-excess ratios can
include class III pre-main sequence stars, main sequence stars and red
giants (clump giants, and RGB stars), all of which have different
intrinsic colors. Our constraint to exclude unreddened stars and bright probes 
should minimize contribution from foreground objects. The proximity of Serpens and Ophiuchus reduces the number of foreground contaminants which must be removed. \citet{ind05} restrict themselves to red clump giant stars
and find they are well-separated from main sequence stars on a $J$ vs. $J-K_s$
color-magnitude diagram for the regions they studied. 
 
\begin{figure}
\epsscale{1.}
\plotone{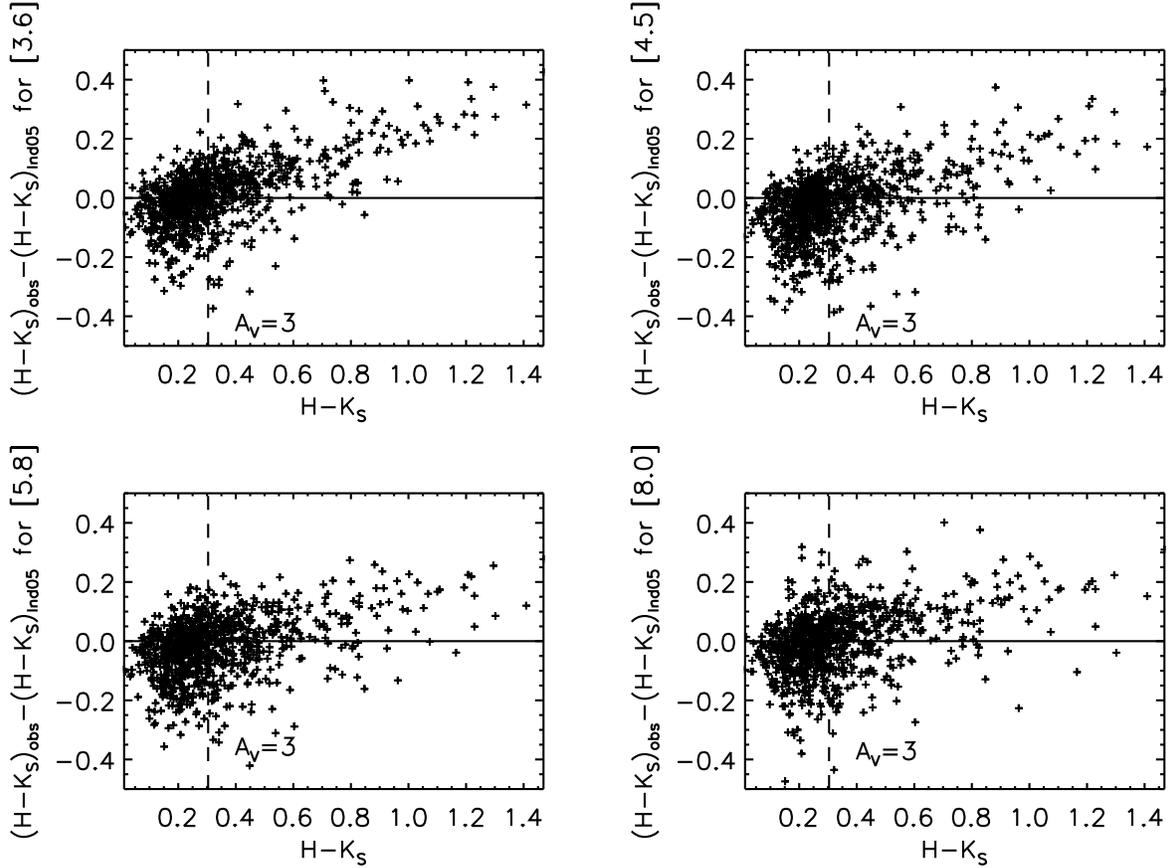}
\caption{Residuals $(H-K_s)_{obs}-(H-K_s)_{Ind05}$ for the non-excess
sources in NGC 2068/71 vs. $H-K_s$. See text for explanation of
residuals.  The solid horizonal line is flat and is the expected slope
if the reddening law \citep{ind05} were a perfect fit to the data. At
larger extinction (larger $H-K_s$) there is a substantial separation from
the solid line, suggesting the appropriate reddening law has a
different slope than that of \citet{ind05}. Vertical dashed line shows the color for a K2 giant \citep{bb88}, taken to be the typical spectral type for a red clump star, extinguished by $A_V$=3 using the extinction law of \citet{rl85}. $A_V$=3 is the boundary for ice mantle growth within a molecular cloud \citep{wh01}.\label{residual}}
\end{figure}

We examine the dependence on extinction probe by calculating residuals which are defined as the
difference between $(H-K_s)_{obs}$ for a given $K_s-[\lambda]$ and the
predicted $(H-K_s)_{Ind05}$ using the \citet{ind05} determination of
$E_{H-K_s}/E_{K_s-\lambda}$. Figure~\ref{residual}, where the residuals are
plotted versus $H-K_s$ color, shows the deviation of our extinction law determination for NGC $2068/71$, using non-excess stars, from that of \citet{ind05} using red clump giants.  As noted in $\S$ 3, the parameterization used by \citet{ind05} to pick out red clump giants will not apply to our cluster lines of sight. As can be seen in Figure~\ref{JvJ-K_2068_off} our extinction probes in  NGC $2068/71$ exhibit varying levels of extinction. If using only red clump giant probes produced a systematically different extinction law than we measured, we would expect a locus of sources to follow the \citet{ind05} law at all four bands. No such locus is observed in Figure~\ref{residual}, suggesting that extinction probes are not the main cause of the systematic difference. Sources with $H-K_s > 0.6$, which are mostly red giants, show a substantial deviation from the \citet{ind05} extinction law. Since both methods referred measurements to the $K_s$-band rather than to
the $K$-band, a correction for different photometric systems is not necessary.

Rederived GLIMPSE color excess
ratios are systematically separated from the color excess ratios for
the five young clusters observed here. While there are differences in
the results derived from the GLIMPSE data using the different methodologies, especially at
3.6 and 4.5\micron, there is still a significant separation between
our color excess ratios and those from \citet{ind05}. This suggests
that the separation is not primarily caused by the differing methods
and/or extinction probes. The systematic separation may instead be due
to differences in the extinction from dust in the molecular clouds and
the diffuse interstellar medium. This possibility will be further
discussed in $\S$~\ref{cloudVism}

\subsubsection{IRAC-derived Extinction Law}
We derive the extinction law $A_{\lambda}/A_{K_s}$ for the five clusters studied here as well as for the selected GLIMPSE data, using the color excess ratios determined above. $A_{\lambda}/A_{K_s}$ is related to the color excess ratios, and the assumed value of $A_H/A_{K_S}$ by:
\begin{equation}
\frac{E_{H-K_S}}{E_{K_S-\lambda}}=\frac{\frac{A_H}{A_{K_S}}-1}{1-\frac{A_{\lambda}}{A_{K_S}}}.
\end{equation}
Since we have used precisely the same methods to establish the extinction law for both data sets, we have removed uncertainties in the comparison that are method-dependent.  In both cases we assume $A_H/A_{Ks}$ = 1.55: values for $A_{\lambda}/A_{K_s}$ are listed in Table~\ref{ext}.  In addition, we also give the extinction law for the $l=284^{\circ}$ off-cloud line of sight as derived by \citet{ind05}. 

Different choices for $A_H/A_{K_s}$
will affect our derived values of $A_{\lambda}/A_{K_s}$. For NGC
$2024/23$, for example, if $A_H/A_{K_s}=1.53$ ($\beta=1.61$) was chosen then
$A_{\lambda}/A_{K_s}$ would be $0.64,0.55,0.51,0.52$ for the four IRAC
bands. If instead $A_H/A_{K_s}$ of 1.73 ($\beta$=1.99) from
\citet{ni06} was chosen then $A_{\lambda}/A_{K_s}$ would be
$0.51,0.38,0.32,0.34$ for the four IRAC bands. The difference in these two determinations of the relative extinction law is larger than the statistical uncertainties in $A_{\lambda}/A_{K_s}$ derived from the uncertainties in the selective color excess ratios, which are $\le0.01$. The uncertainties reported in Table~\ref{ext} include the uncertainties in the measured color excess ratios but do not include the uncertainty in $A_H/A_{K_s}$.

\begin{figure}
\epsscale{.7}
\plotone{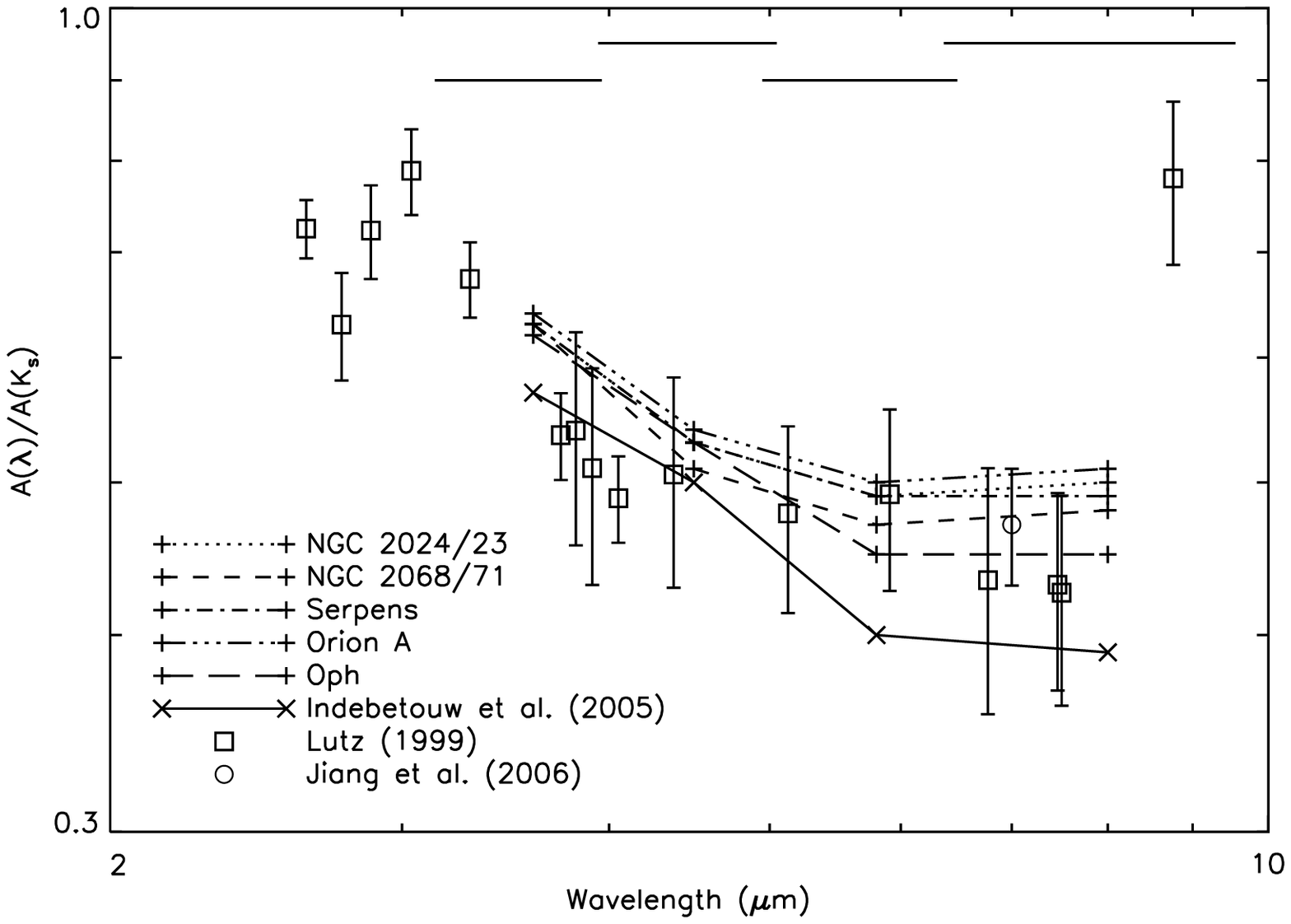}
\plotone{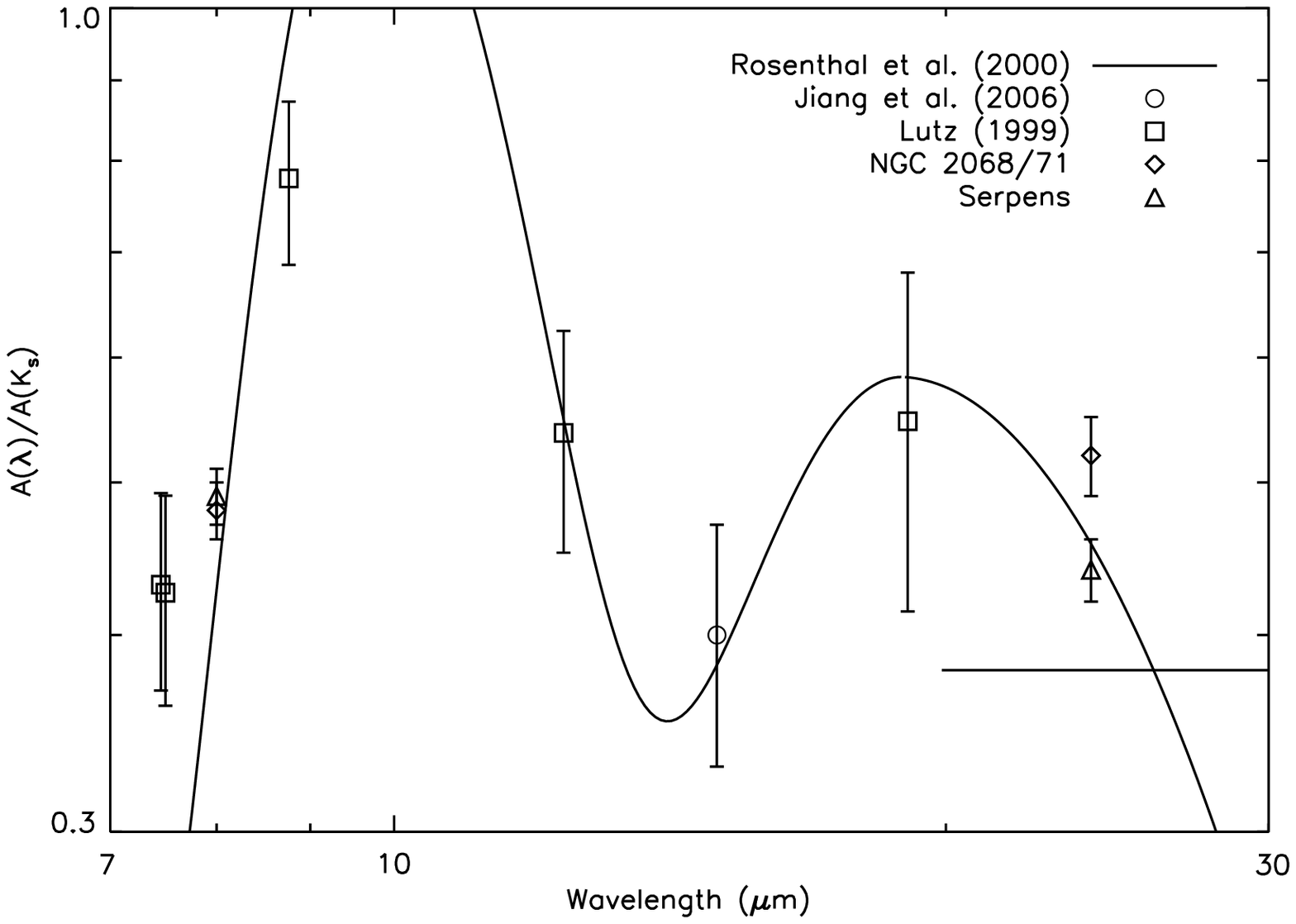}
\caption{Extinction in IRAC bands versus wavelength for each of NGC
$2023/24$, NGC $2068/71$, Serpens, Ophiuchus, Orion A and the extinction law rederived here for the $l=284^{\circ}$ line of sight from \citet{ind05}. The squares are the extinction law toward the galactic
center as measured by \citet{lut99} using hydrogen recombination
lines. On the top is the extinction law from $1-10\micron$, while on
the bottom is the extinction law from $7-30\micron$. The extinction law
measured at $7\micron$ and $15\micron$ by \citet{ji06} are labeled
with circles in the two plots. On the bottom, the extinction law between 7
and $24\micron$ is shown for NGC 2068/71, Serpens and as measured by
\citet{lut99} with diamonds, triangles and squares respectively. The
line in the second plot is the extinction curve derived by
\citet{ros00} using $H_2$ line ratios observed towards OMC-1. Horizontal solid lines represent the bandpasses for the IRAC and MIPS filters in the top and bottom plots repectively.\label{ext_wav}}
\end{figure}

The values of $A_{\lambda}/A_{K_s}$ for each of the five regions observed here, subject to the assumption of $A_H/A_{K_s}=1.55$, are plotted in Figure~\ref{ext_wav}. The rederived extinction law for the $l=284^{\circ}$ off-cloud line of sight from \citet{ind05} is also plotted in Figure~\ref{ext_wav}. Extinction law measurements by \citet{lut99}, included as squares in Figure~\ref{ext_wav}, are more consistent with \citet{ind05}, although some of their measurments overlap with our results. The measurement at $7\micron$ from \citet{ji06} is marked with a circle in Figure~\ref{ext_wav}, and the spread in their results for many different sight lines encompass both our results and the results from \citet{ind05}. Our extinction law deviates from a power law, as is also seen in these previous studies.

\subsubsection{Comparison with ISO-Derived Extinction Laws from $3-8\micron$}\label{iso}

\citet{lut99} compared the observed to expected fluxes for 15 hydrogen
recombination lines toward the galactic center from ISO data and from
this determined the extinction ratios at the wavelengths of the
emission lines. Theoretical estimates of the expected fluxes assumed
T$_e$ = 7000K, n$_e$=3000 cm$^{-3}$, case B recombination (which they
verify from their observations) and refer directly to the calculations
by \citet{st95}. The recombination line results also depend on the
assumed helium abundance, [HeII]/[HeIII], and the assumption that line
emissions emanate from the same volume. A crucial part of their
analysis is the scaling of their relative extinction law: they deduce
$A_\lambda$/$A_{2.626}$ (Br$\beta$ at 2.626$\mu$m is the shortest-wave
emission line they observe) and use a $\lambda^{-1.75}$ scaling law to
refer to $K$-band. We will use a $\lambda^{-1.65}$ law which corresponds to our assumed $A_H/A_{Ks}$ = 1.55, allowing for a more direct comparison with our values of $A_{\lambda}/A_{K_s}$ (see discussion below).  Numerical data on $A_{\lambda}$/$A_V$ were provided
courtesy of D. Lutz: here we deduce from these data
$A_\lambda$/$A_{2.626}$, and using a factor of
$A_{K_s}$/$A_{2.626}$=1.38 (assuming the power law index $\beta$=1.65),
deduce $A_\lambda$/$A_{K_s}$ for their data. Since we use 2MASS
$JHK_s$ data, we have taken the wavelength for $K_s$-band to be
2.16$\mu$m.  Power law indices of 1.61 and 1.8 lead to at most a 2\%
error in the extinction ratio.  This is considerably more certain than
variations in $A_K/A_V$, which can be as much as 30\% \citep{gl99}, so
we choose not to compute $A_\lambda/A_V$ for our data.

\citet{ji06} combined ISO observations at 7 and 15$\micron$ of the
inner galactic plane with 2MASS and DENIS near-infrared observation to
determine the mid-infrared extinction law and its variation along
different sightlines. They assumed a constant value of $(J-K_s)_o$,
$(K_s-[7])_o$ and $(K_s-[15])_o$ for early AGB and RGB-tip stars and
measured the color excess of these stars along various
sightlines. Assuming $A_J/A_K$ from \citet{rl85} they find
$A_7/A_K=0.47\pm0.04$ and $A_{15}/A_K=0.40\pm0.07$ (marked as circles
in figure~\ref{ext_wav}). Although \citet{rl85} did not measure the
extinction law with the 2MASS photometric system, \citet{ind05} found
$A_J/A_{K_s}$ in the 2MASS photometric system to be similar to the value of
$A_J/A_K$ measured by \citet{rl85}. The results from \citet{ji06} do
not change significantly when using $A_J/A_{K_s}$ from \citet{ind05}
as opposed to $A_J/A_K$ from \citet{rl85}. At 7$\micron$ they find the
extinction law measurements have a Gaussian distribution whose one
sigma size encompasses the measurements presented here, as well as
those of \citet{ind05} and \citet{lut99}. There is more uncertainty at
$15\micron$ but there also appears to be evidence for a Gaussian
distribution of extinction values, $A_{15}/A_K$.

We chose $A_H/A_{K_s}$ from \citet{ind05} to provide a direct comparison of all the mid-infrared selective extinction law measurements discussed here. The extinction law measurements by \citet{ji06} use a similar normalization to our assumed $A_H/A_{Ks}$, and the measurements of \citet{lut99} were normalized using a power law form that is consistent with $A_H/A_{K_s}=1.55$. Different assumptions of the near-infrared extinction law will affect our derived $A_{\lambda}/A_{K_s}$ as discussed earlier, but we choose a single consistent normalization to better compare the various determinations of the mid-infrared extinction law.

\subsubsection{Molecular Cloud vs. Diffuse Interstellar Extinction.}\label{cloudVism}
A plausible explanation for the separation between our extinction law and
the extinction law of \citet{ind05} is a variation in the extinction
law between that through the diffuse ISM and through molecular clouds. Variation in dust optical properties
between molecular cloud dust and that in the ISM has been invoked by \citet{hu07}, who, from observations of cloud cores obtained as part of the c2d {\it Spitzer} Legacy program, have also found 
extinction law slopes (color excess ratios) which deviate from those of \citet{ind05}.

The cold ISM includes diffuse interstellar clouds ($A_V <$ 0.1), poorly shielded regions of molecular clouds ($A_V <$ 3), and dense molecular cloud cores ($A_V >$ 3). \citet{wh01} and references therein, have explored grain growth in dense clouds, and have reported differing extinction laws for the diffuse ISM and dense clouds.  For reference, $A_V$=3 is shown as a vertical dashed line in  Figure~\ref{residual}: most of the deviation of our results from the \citet{ind05} extinction law occurs at higher extinctions. Changes in the extinction law originating in these regions are likely due to grain growth, either by coagulation of small grains or the growth of icy mantles, both of which will affect the continuum level of extinction. Small grains are thought to coagulate to form larger grains (e.g. $\rho$ Oph, \citet{wh01}): grain emissivity is dependent on the optical constituent properties, and for the
same constituents is proportional to a power(powers) of {\it a}/${\lambda}$ where {\it a} is the grain size. Grain growth will not only remove small grains as potential absorbers, but also as emitters. The lack of emission from small grains has been observed along lines of sight with high column densities \citep{ste03,bur03, cam05} suggesting grain growth leading to porous grains with enhanced emissivity within these clouds. \citet{pend06} have found a decrease in the depth of the 9.7 $\mu$m silicate feature with increasing $A_V$, indicating silicate grain growth.  $A_V = $ 3 is the approximate extinction threshold for the detection of ice mantles, and in the Taurus Dark Cloud, icy mantle growth is implicated \citep{wh01}. Dust in molecular cloud cores is protected both from the UV interstellar radiation field and from cosmic ray processing; those dust grains may therefore have larger mantles consisting of water and organic ices \citep{gibb04}. Absorption features between 3 and 8\micron\ due to ices are often seen along lines of sight intersecting dense molecular clouds \citep{kn05,gibb04} while they are not seen in the diffuse ISM \citep{wh97}. On the other hand, grain processing and fragmentation because of radiative shock propagation, and turbulence, may also be involved in grain evolution in both molecular clouds and in the diffuse ISM. 

Our extinction law for the young clusters studied here derives mainly from dust associated with their molecular clouds, while that of \citet{ind05} derives primarily from diffuse interstellar dust, especially for the $l=284^{\circ}$ off-cloud line of sight. There is a greater contribution from diffuse interstellar dust for the closest clusters in our sample, Ophiuchus and Serpens, partially because they are in the inner galaxy, and partially because the extinction probes (not excluded by the $JHK$ brightness cutoff) lie well behind the cluster, so that their light passes through more of the ISM before reaching the cluster. The extinction curve toward Ophiuchus has been termed anomalous: measurements of the UV/optical extinction curve and linear polarization \citep{gre92,kim96} toward Ophiuchus indicate a depletion of small grains and the presence of large grains within the molecular cloud.  Studies of other dark clouds have produced similarly anomalous extinction as compared with the ISM: we are unaware of any comprehensive study comparing optical/UV extinction in the dark clouds studied here. \citet{tx99} observe H$_2$O and CO ices in the Taurus and Ophiuchus dark clouds, and concur with previous studies that different physical conditions obtain in those two clouds.  Although column densities of the H$_2$O ices and the visual extinction are correlated in each case, the regression line is different.  There are several Ophiuchus sources which follow the Taurus regression line rather than that for Ophiuchus, suggesting physical properties of the grains can also vary within the cloud.  \citet{sch05} and \citet{dup03} measure the emissivity spectral index $\beta$ ($\kappa_\nu \propto \nu^{\beta}$) in the far-infrared and find variations between Orion, Ophiuchus and Serpens. This indicates that the grain properties differ between these three clouds, but the anticorrelation between spectral index and temperature within a cloud suggests that there can be significant variations in grain properties along different lines of sight through a single cloud. Consequently, we cannot determine here the extent to which the molecular cloud mid-IR extinction toward Ophiuchus is anomalous compared with other clouds in our sample.

In contrast to our clusters, all within 500 pc of the Sun,  the lines of sight for the ISO measurement may include a larger contribution from a mixture of extinction from the diffuse ISM and molecular clouds. The \citet{lut99} extinction law to the galactic center ionized gas is dominated by dust in the diffuse ISM along the line of sight, although some of the extinction is due to molecular cloud material \citep{lut96,wh97}. Broad ice absorption features at 3 and 6\micron\ are seen toward Sgr A* and are attributed to foreground molecular cloud material \citep{chi00}.  The \citet{ji06} extinction law is for a mix of diffuse ISM and dense molecular cloud dust.  Given the uncertainties in the derived laws of \citet{ji06} and \citet{lut99}, we cannot distinguish whether any agreement, or lack of agreement with either our law or that of \citet{ind05} depends on whether the material causing extinction is part of the diffuse ISM or a dense cloud.
 
The differing dust properties between diffuse ISM and dense molecular clouds could lead to differences in the extinction law.  We directly measure $E_{H-K_s}/E_{K_s-[\lambda]}$, which relates
$A_{[\lambda]}/A_{K_s}$ and the assumed value of $A_H/A_{K_s}$. A variation in the near-infrared extinction law, the mid-infrared extinction law, or both, could lead to the observed systematic separation in the selective color excess ratios between our observations through molecular clouds  and \citet{ind05}'s observations through the diffuse ISM. If a
larger value of $A_H/A_{K_s}$ was more appropriate for the extinction
law toward molecular clouds, then $A_{[\lambda]}/A_{K_s}$ would be
smaller and the {\it mid-infrared} portion of the extinction law
deduced by \citet{ind05} and us could be made to agree. Values of
$A_H/A_{K_s}$ between 1.6 and 1.65, depending on the region, would
bring our mid-infrared extinction law into agreement with the average extinction law reported by
\citet{ind05}, assuming $A_H/A_{K_s}=1.55$ still applied to the
\citet{ind05} extinction law. It may also be the case that the
near-infrared extinction law does not appreciably vary between the
diffuse ISM and molecular clouds and there is instead a variation in
the mid-infrared extinction law. We now examine the near- and mid-infrared data separately in order to attempt to break this degeneracy.

The possibility of variations in $A_H/A_{K_s}$ is suggested by recent observations. Although \citet{car89} and \citet{mw90} found the near-infrared
extinction law to be universal, more recent studies have produced
mixed results. One of the difficulties in evaluating the variation in
the near-infrared extinction law along different lines of sight is
that different photometric systems (e.g. $K$ vs. $K_s$) can have a
substantial effect on the measured extinction law
\citep{klb98}. \citet{klb98} and \citet{gk01} observed the nearby
star-forming regions $\rho$ Oph and Cha I, respectively, and
determined the near-infrared reddening law using the same technique,
but with different photometric systems, and find a substantial
difference in the reddening law between these two lines of
sight. \citet{na06} reobserved $\rho$ Oph and Cha I with one
photometric system and used the same technique, the same off-cloud
regions, and the same limiting magnitudes as the previous studies by
\citet{klb98} and \citet{gk01} but found no variation in the
near-infrared reddening law between these two molecular
clouds. \citet{ni06} measured the extinction law toward the galactic
center using the same photometric system as \citet{na06} and their
results are consistent with the reddening law being universal between
the Galactic Center and the $\rho$ Oph and Cha I star-forming
regions. Evidence for a variation in the near-infrared reddening law
comes from \citet{rgk02} who observed the Coalsack Globule 2 using the
same photometric system as \citet{gk01}, and used the same technique
to determine the reddening law, but found a significant difference
between the reddening laws along these two lines of sight.

To isolate the influence of the near-infrared extinction law on our measured color-excess ratios we have measured $E_{H-K_{s}}/E_{K_s-J}$ in a manner similar to the previous color-excess ratios. Since we did not require a J band detection for the Orion A and Ophiuchus regions, we excluded these regions. NGC 2024/23, NGC 2068/71 and Serpens have color excess ratios of $-0.333\pm0.004, -0.335\pm0.003$ and $-0.328\pm0.004$. The {\it l}=$284^{\circ}$ off-cloud region has a color excess ratio of $-0.326\pm0.004$. No significant difference exists in the near infrared color excess ratios among these lines of sight within uncertainties, indicating that there is no significant variation in the near-infrared extinction law through these three molecular clouds and through the interstellar medium.

We examine the mid-infrared portion of the extinction law by forming color-excess ratios that include only IRAC bands from the already measured color excess ratios. From the definition of the color excess ratio
\begin{equation}
R_{\lambda}=\frac{E_{H-K_s}}{E_{K_s-[\lambda]}}=\frac{A_H-A_{K_s}}{A_{K_s}-A_{\lambda}}
\end{equation}
\noindent we form the following quantity:
\begin{equation}
\frac{\frac{1}{R_{\lambda2}}-\frac{1}{R_{\lambda1}}}{\frac{1}{R_{\lambda4}}-\frac{1}{R_{\lambda3}}}=\frac{A_{\lambda1}-A_{\lambda2}}{A_{\lambda3}-A_{\lambda4}}.
\end{equation}
By choosing $\lambda1=3.6\micron, \lambda2=\lambda3=4.5\micron$ and $\lambda4=5.8\micron$ we have a measure of the IRAC-only color-excess ratio. By combining color-excess ratios rather than using the extinction law measurements, $A_{\lambda}/A_{K_s}$, we reduce the uncertainty in the IRAC-only color-excess ratio. The results are listed in Table~\ref{irac_ratio} for all five young clusters as well as for the $l$=$284^{\circ}$ off-cloud line of sight. While the uncertainties are large, there is clear evidence for a separation between the extinction law measured toward the molecular clouds and the extinction law for only the diffuse ISM. The tabulated values of this ratio demonstrate that the variations between our color-excess ratios and those of \citet{ind05} cannot be explained solely by a change in the value of $A_H/A_{K_S}$, and that the form of the interstellar extinction law must differ between the diffuse ISM, Ophiuchus, and the remaining four molecular clouds. It is also apparent that the Ophiuchus cloud reddening law differs from the reddening law for the other four young clusters. 

\subsection{The Extinction Law beyond $8\micron$}\label{24micron}
\begin{figure}
\epsscale{.7}
\plotone{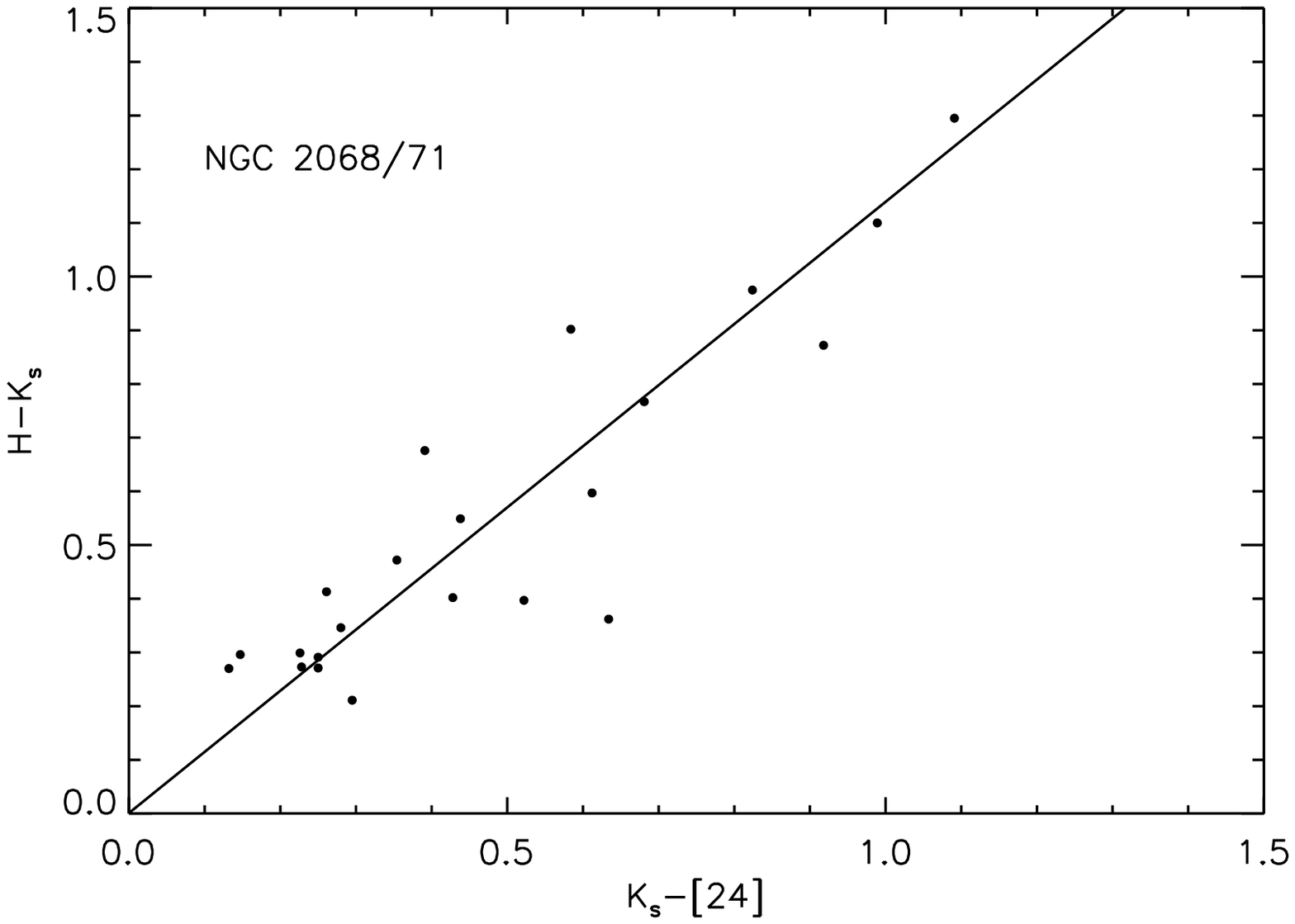}
\plotone{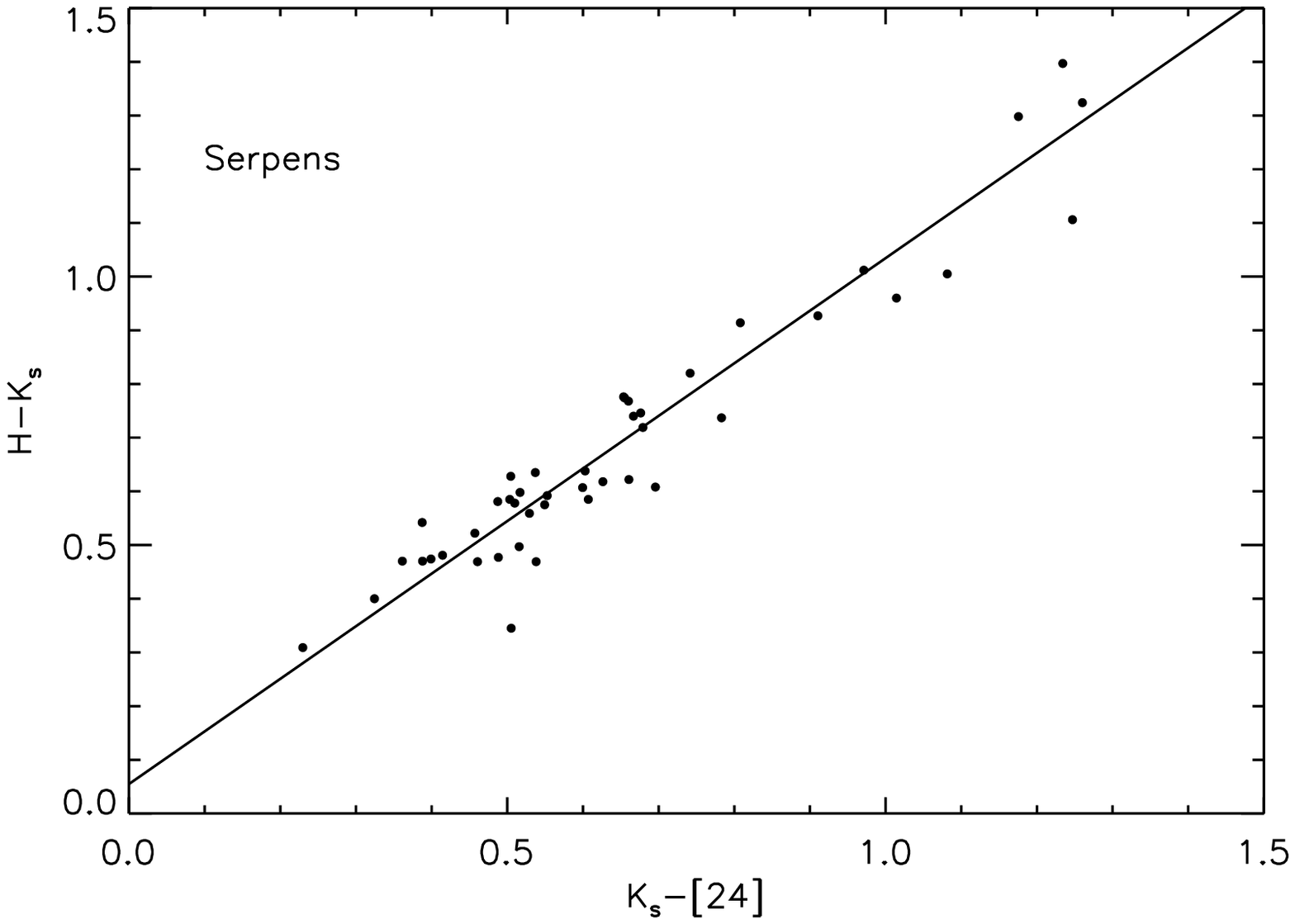}
\caption{2MASS and MIPS 24$\micron$ color-color diagrams for NGC $2068/71$ (top) and Serpens (bottom). Line is the best fit to the data.\label{24_red}}
\end{figure}

Extending our analysis out to 24\micron\ we measure $E_{H-K}/E_{K-[24]}=0.98\pm0.04$ and
$1.14\pm0.06$, $A_{24}/A_{K_s}=0.44\pm0.02$ and $0.52\pm0.03$, assuming $A_H/A_{K_s}=1.55$, for Serpens and NGC 2068/71 respectively. The large
uncertainty for both regions' color excess ratios reflects the small number of sources used in the
fitting procedure. Figure ~\ref{24_red} shows the fit to the reddening
bands at 24$\micron$, similar to Figures~\ref{ngc2024_red} -
~\ref{oriona_red}. Figure ~\ref{ext_wav} illustrates that our
24$\micron$ determinations are in reasonable accord with those of
\citet{lut99} shown as squares in Figure~\ref{ext_wav}, given the
large uncertainties, as well as with the measurement at 15$\micron$ by
\citet{ji06}, $A_{15}/A_K=0.15\pm0.07$, shown as a circle in
Figure~\ref{ext_wav}. \citet{hu07} also measured the extinction law at
$24\micron$ and found it to be consistent with our results.

\citet{ros00} measured the extinction law from $2-30\micron$ toward
OMC-1 using the relative intensities of $H_2$ rotation and
ro-vibration lines. They fit a functional form assuming a power law
for $\lambda<6.5\micron$ and included the silicate features at
$9.7\micron$ and $18\micron$, as well as a water ice feature at
$3.1\micron$. The strength of the $9.7\micron$ silicate feature was
determined from a fit to their observations, while the relative
strength of the $18\micron$ and $9.7\micron$ silicate feature was
taken to be $A_{18}/A_{9.7}=0.44$, an average of previous results from
\citet{dr84}, \citet{pp85} and \citet{vk88}. The shape of the 9.7\micron\ feature was chosen to match the Trapezium emissivity profile, which is more appropriate for extinction due to dense molecular clouds rather than the narrower $\mu$ Cephei profile \citep{bow98}. While the assumption of a
power law from $3-8\micron$ is clearly not valid, the extinction law
from $10-30\micron$, as measured here and by \citet{lut99} and
\citet{ji06}, is very well fit by a law dominated by silicate emission with the  adopted strengths of the two silicate features given by \citet{ros00}, as can be
seen in Figure~\ref{ext_wav}. The $9.7\micron$ silicate feature has been found to vary along different lines of sight \citep{bow98}, particularly on the long-wavelength side of the feature, although there is insufficient data here to draw any conclusions about the universality of the extinction law from $8-30\micron$.

\section{Conclusion}
We measured the extinction law from $3 - 8\micron$ toward five nearby
star-forming regions and found it to be relatively internally
consistent among different cluster-forming molecular clouds although there is evidence that the Ophiuchus extinction law deviates from the other four regions at 5.8 and 8\micron. The
extinction laws we present here differ systematically from the
extinction at IRAC wavelengths derived by \citet{ind05}, as seen in
both the color excess ratios and the relative
extinction ratios. To exclude any possibility that this difference could be due to the differing
methodologies used to derive the extinction law, we have also analysed data provided by R. Indebetouw using our methods, and continue to find systematic variation.  The result could reflect a
physical difference in the extinction law between the diffuse ISM and
molecular clouds, or the differing extinction probes utilized in the two studies. Our study of the probes leads us to believe that the major cause of the discrepancy is variation in the diffuse ISM vs. molecular cloud extinction laws. The observed difference in the extinction law towards Ophiuchus could be caused by diffuse ISM behind the molecular cloud that the extinction probes must pass through or differing dust properties between the Ophiuchus cloud and the other four clouds studied.

The extinction law derived by \citet{lut99} agrees more closely with the re-derived \citet{ind05} extinction law at wavelengths $\leq 3\mu$m. At 3$\mu$m $\leq \lambda \leq 8\mu$m, the opposite is true. At $7\micron$ the relative extinction measured by \citet{ji06} agrees with the cluster results. The extension of our extinction law to
$24\micron$ was also determined from analysis of {\it Spitzer} images
of Serpens and NGC 2068/71, but further observations at this
wavelength are required to provide statistically significant
conclusions. However, our results at $24\micron$, along with the results of
\citet{lut99} and \citet{ji06}, match well the functional form of
\citet{ros00} which includes extinction dominated by silicate features at 9.7
and 18$\micron$. Future measurements of the extinction law for
$10-30\micron$ will help to better constrain this fit, as well as to
examine any variation among different lines of sight.

\clearpage

\begin{deluxetable}{lcccc}
\tablewidth{0pt}
\tablecaption{Observing Information}
\tablehead{\colhead{Region}&\colhead{GTO Program IDs}&\colhead{Pipeline Version}&\colhead{Magnitude Zero Points\tablenotemark{a}}&\colhead{Frame Time\tablenotemark{b}}}
\startdata
NGC $2024/23$&43&S9.5.0&21.93,21.26,19.08,19.44&12\\
NGC $2068/71$&43&S9.5.0&21.93,21.26,19.08,19.44&12\\
Orion A&43&S9.5.0&21.93,21.26,19.08,19.44&12\\
Serpens&6,174,177&S11.4&21.99,21.26,19.07,19.44&12\\
Ophiuchus&6,174,177&S11.4&21.99,21.26,19.07,19.44&12\\
\enddata
\tablenotetext{b}{HDR mode includes the time spent on both the long and short exposure. The effective exposure time after mosaicing the multiple maps of each region (two for NGC 2024/23, NGC2068/71 and Orion A, four for Serpens and Ophiuchus) will be longer than this frame time.}
\tablenotemark{a}{For an exposure time of 10.4 seconds and object, inner and outer sky apertures of 2,2 and 6 pixels respectively.}
\end{deluxetable}

\begin{deluxetable}{lcccc}
\tablewidth{0pt}
\tablecaption{Selective Color Excess Ratios\label{slopes}}
\tablehead{\colhead{Region}&\colhead{$E_{H-K_s}/E_{K_s-[3.6]}$}&\colhead{$E_{H-K_s}/E_{K_s-[4.5]}$}&\colhead{$E_{H-K_s}/E_{K_s-[5.8]}$}&\colhead{$E_{H-K_s}/E_{K_s-[8.0]}$}}
\startdata
NGC $2024/23$&$1.49\pm0.02(0.08)$&$1.17\pm0.02(0.06)$&$1.08\pm0.01(0.06)$&$1.11\pm0.02(0.07)$\\
NGC $2068/71$&$1.49\pm0.02(0.07)$&$1.13\pm0.01(0.05)$&$1.03\pm0.01(0.05)$&$1.07\pm0.01(0.05)$\\
Serpens&$1.49\pm0.02(0.08)$&$1.17\pm0.02(0.06)$&$1.08\pm0.01(0.05)$&$1.07\pm0.01(0.06)$\\
Orion A&$1.51\pm0.01(0.04)$&$1.197\pm0.008(0.03)$&$1.109\pm0.007(0.03)$&$1.116\pm0.008(0.03)$\\
Ophiuchus&$1.46\pm0.02(0.07)$&$1.17\pm0.01(0.05)$&$1.01\pm0.01(0.05)$&$1.01\pm0.01(0.05)$\\
Ind$05$data&$1.27\pm0.02(0.09)$&$1.11\pm0.02(0.08)$&$0.92\pm0.02(0.06)$&$0.90\pm0.01(0.06)$\\
\tableline
Ind$05$paper&$1.17\pm0.07$&$1.0\pm0.03$&$0.92\pm0.03$&$0.92\pm0.04$\\
\enddata
\tablecomments{Values ``Ind05 paper" are derived from color excess rations from \citet{ind05} with their uncertainties propogated, and ``Ind05 data" are derived from the data R. Indebetouw kindly provided.  Both are for the $l=284^{\circ}$ off-cloud line of sight data.  Ind05 data were analyzed in the same way that our data were analyzed, in order to enable direct comparison with our results. Our uncertainties are for a $68\%$ confidence level with uncertainties for a $99.99\%$ confidence interval in parenthesis.}
\end{deluxetable}

\begin{deluxetable}{lcccccc}
\tablewidth{0pt}
\tablecaption{Relative Extinction\label{ext}}
\tablehead{\colhead{Region}&\colhead{$A_{[3.6]}/A_{K_s}$}&\colhead{$A_{[4.5]}/A_{K_s}$}&\colhead{$A_{[5.8]}/A_{K_s}$}&\colhead{$A_{[8.0]}/A_{K_s}$}}

\startdata NGC $2024/23$&$0.632\pm0.005$&$0.53\pm0.01$&$0.49\pm0.01$&$0.50\pm0.01$\\
NGC $2068/71$&$0.632\pm0.005$&$0.51\pm0.01$&$0.47\pm0.01$&$0.48\pm0.01$\\
Serpens&$0.630\pm0.005$&$0.53\pm0.01$&$0.49\pm0.01$&$0.49\pm0.01$\\
Orion A&$0.636\pm0.003$&$0.540\pm0.003$&$0.504\pm0.003$&$0.506\pm0.003$\\
Ophiuchus&$0.623\pm0.005$&$0.53\pm0.01$&$0.45\pm0.01$&$0.45\pm0.01$\\
Ind$05$ data&$0.57\pm0.01$&$0.50\pm0.01$&$0.40\pm0.01$&$0.39\pm0.01$\\
\tableline
Ind$05$ paper&$0.57\pm0.05$&$0.43\pm0.07$&$0.41\pm0.07$&$0.37\pm0.07$\\
\enddata
\tablecomments{Quoted values taken from \citet{ind05} ("Ind05 paper") are for the $l=284^{\circ}$ off-cloud line of sight. The values of $A_{\lambda}/A_{K_s}$ were derived from the measured selective color excess ratios $E_{H-K_s}/E_{K_s-\lambda}$ for the five young clusters and the rederived color excess ratios for the $l=284^{\circ}$ line of sight assuming $A_H/A_{K_s}=1.55\pm0.08$ \citep{ind05}. Uncertainties on $A_{\lambda}/A_{K_s}$ derived here do not include the uncertainty in $A_H/A_{K_s}$. Uncertainties on $A_{\lambda}/A_{K_s}$ taken from \citet{ind05} (labeled Ind05 paper) do include the uncertainty in $A_H/A_{K_s}$.}
\end{deluxetable}

\begin{deluxetable}{lc}
\tablewidth{0pt}
\tablecaption{IRAC-only Color Excess Ratio}
\tablehead{\colhead{Region}&\colhead{$E_{[3.6]-[4.5]}/E_{[4.5]-[5.8]}$}}

\startdata
NGC $2024/23$&$2.58\pm0.66$\\
NGC $2068/71$&$2.49\pm0.56$\\
Serpens&$2.58\pm0.66$\\
Orion A&$2.61\pm0.33$\\
Ophiuchus&$1.25\pm0.13$\\
\tableline
Ind$05$ data&$0.61\pm0.14$\\
\enddata
\tablecomments{Color Excess ratio derived using $E_{H-K_s}/E_{K_s-[\lambda]}$. See text for more details on this calculation.\label{irac_ratio}}
\end{deluxetable}

\clearpage

\begin{thebibliography}{}
\bibitem[Allen et al.(2004)]{all04} Allen, L.E., et al. 2004, \apjs, 154, 363
\bibitem[Allen et al.(2007a)]{allen07} Allen, L.E., et al. 2007a in Protostars and Planets V, ed B. Reipurth, D. Jewitt, \& K. Keil (Tucson: Univ. Arizona Press), 361
\bibitem[Allen et al.(2007b)]{all07} Allen, L.E., et al. 2007b in prep
\bibitem[Bertoldi et al.(1999)]{ber99} Bertoldi, F., Timmermann, R., Rosenthal, D., Dropatz, S. \& Wright, C.M. 1999, \aap, 346, 267
\bibitem[Bessell \& Brett(1988)]{bb88} Bessel, M.S. \& Brett, J.M. 1988, \pasp, 100, 1134
\bibitem[Bowey et al.(1998)]{bow98} Bowey, J.E., Adamson, A.T. \& Whittet, D.C.B. 1998, \mnras, 298, 131
\bibitem[Cambresy et al.(2005)]{cam05} Cambresy, L., Jarrett, T.H., \& Beichman, C.A. 2005, \aap, 435, 131
\bibitem[Cardelli et al.(1989)]{car89} Cardelli, J.A., Clayton, G.C. \& Mathis, J.S. 1989, \apj 345, 245
\bibitem[Chiar et al.(2000)]{chi00} Chiar, J.E., Tielens, A.G.G.M., Whitter, D.C.B., Shuttle, W.A., Boogert, A.C.A., Lutz, D., van Dishoeck, E.F. \& Bernsteain, M.P. 2000, \apjs, 154, 363
\bibitem[Cohen(1993)]{co93} Cohen, M. 1993, \aj 105, 1860
\bibitem[del Burgo et al.(2003)]{bur03} del Burgo, C., Laurejis, R.J., \'Abrah\'am, P. \& Kiss, Cs. 2003, \mnras, 346, 403
\bibitem[Draine(1989)]{dr89} Draine, B.T. 1989 Proc. 22nd Eslab Symposium on Infrared Spectroscopy, ESA SP-290, 93.
%\bibitem[Draine(2003)]{dra03} Draine, B.T. 2003 \araa, 41, 241
\bibitem[Draine \& Lee(1984)]{dr84} Draine, B.T. \& Lee H. 1984, \apj, 285, 89
\bibitem[Dupac et al.(2003)]{dup03} Dupac, X., et al., 2003, \aap, 404, L11
\bibitem[Evans et al.(2003)]{ev03} Evans, N.J., II, et al. 2003, \pasp, 115, 965
\bibitem[Fazio et al.(2004)]{faz04} Fazio, G.G., et al. 2004, \apjs, 154, 10 
\bibitem[Gibb et al.(2004)]{gibb04} Gibb, E.L., Whittet, D.C.B., Boogert, A.C.A. \& Tielens, A.G.G.M. 2004, \apjs, 151, 35
\bibitem[Giles(1977)]{gil77} Giles, K. 1977, \mnras, 180, 57P
\bibitem[Glass(1999)]{gl99} Glass, I. 1999, Handbook of Infrared Astronomy (Cambridge University Press)
\bibitem[Gomez \& Kenyon(2001)]{gk01} Gomez, M. \& Kenyon, S. 2001, \aj, 121, 974
%\bibitem[Gordon et al.(2005)]{gor05} Gordon, K., et al. 2005, \pasp, 117, 503
\bibitem[Green et al.(1992)]{gre92} Green, J.C., Snow, T.P., Cook, T.A, Cash, W.C. \& Poplawski, O. 1992, \apj, 395, 289
\bibitem[Gutermuth(2005)]{gu05} Gutermuth, R.A. 2005, PhD thesis.
\bibitem[Gutermuth et al.(2004)]{gut04} Gutermuth, R.A., et al. 2004 \apj, 154, 374
\bibitem[Huard et al.(2007)]{hu07} Huard, T., et al. 2007 in prep
\bibitem[Indebetouw et al.(2005)]{ind05} Indebetouw, R., et al. 2005, \apj, 619,931
\bibitem[Jiang et al.(2006)]{ji06} Jiang, B.W., Gao, J., Omont, A., Schuller, F. \& Simon, G. 2006, \aap, 446, 551
%\bibitem[Kaas(1999)]{ka99} Kaas 1999, \aj 118, 558
\bibitem[Kenyon, Lada \& Barsony(1998)]{klb98} Kenyon, S., Lada, E. \& Barsony, M., 1998, \aj, 115, 252
\bibitem[Kim \& Martin(1996)]{kim96} Kim, S.-H. \& Martin, P.G. 1996, \apj, 462, 296
\bibitem[Knez et al.(2005)]{kn05} Knez, C., et al. 2005, \apj, 635, 145
\bibitem[Lada et al.(2006)]{lada06} Lada, C.J., et al. 2006, \aj, 131, 1574
\bibitem[Lada et al.(2007)]{lada07} Lada, C.J., Alves, J.F. \& Lombardi, M. 2007, in Protostars and Planets V, ed. B Reipurth, D. Jewitt, \& K. Keil (Tucson: Univ. Arizona Press), 3
%\bibitem[Landsman(1993)]{lan93} Landsman, W.B. 1993 in Astronomical Data Analysis Software and Systems II, A.S.P. Conference Series, Vol. 52, ed. R.J. Hanisch, R.J.V. Brissenden, and Jeannette Barnes, p.246
\bibitem[Lombardi et al.(2006)]{lom06} Lombardi, M., Alves, J.F. \& Lada, C.J. 2006, \aap, 454, 781
\bibitem[L\'opez-Corredoira et al.(2002)]{lop02} L\'opez-Corredoira, M. Cabrera-Lavers, A., Garz\'on, F. \& Hammersley, P.L. 2002, \aap, 394, 883
\bibitem[Lutz et al.(1996)]{lut96} Lutz, D., et al. 1996, \aap, 315, L269
\bibitem[Lutz(1999)]{lut99} Lutz, D. 1999, in The Universe as Seen by {\it ISO}, ed. P. Cox \& M.F.Kessler (ESA SP-427;Noordwijk:ESA), 623
\bibitem[Martin \& Whittet(1990)]{mw90} Martin, P.G. \& Whittet, D.C.B. 1990, \apj, 357, 113
\bibitem[Megeath et al.(2004)]{mea04}  Megeath, S. T., Gutermuth, R. A., 
Allen, L.E., Pipher, J. L., Myers, P. C. \& Fazio, G. G., 2004, \apjs, 154, 367
%\bibitem[Moore et al.(2005)]{mo05} Moore, T., Lumsden, S., Ridge, N., Puxley, P., 2005, \mnras, 359, 589
\bibitem[Muzerolle et al.(2004)]{muz04} Muzerolle, J., et al. 2004, \apjs, 154, 379
\bibitem[Muzerolle et al.(2007)]{muz07} Muzerolle, J., et al. 2007 in prep
\bibitem[Nishiyama et al.(2006)]{ni06} Nishiyama, S., et al. 2006, \apj, 638, 839
\bibitem[Naoi et al.(2006)]{na06} Naoi, T., et al. 2006, \apj, 640, 373
\bibitem[Pegourie \& Papoular(1985)]{pp85} Pegourie, B. \& Papoular, R.  1985, \aap, 142, 451
\bibitem[Pendleton et al.(2006)]{pend06} Pendleton, Y., et al. 2006, \baas, 208.4915
\bibitem[Press et al.(1992)]{nr92} Press, W.H., Teukolsky, S.A., Vetterling, W.T. \& Flannery, B.P. 1992 ``Numerical Recipes in C: The Art of Scientific Computing'', (Cambridge:Cambridge University Press)
\bibitem[Racca et al.(2002)]{rgk02} Racca, G., Gomez, M. \& Kenyon, S., 2002 \aj, 124, 2178
\bibitem[Reach et al.(2005)]{Reach05} Reach, W.T., et al. 2005, \pasp, 117, 978.
\bibitem[Rieke et al.(2004)]{ri04} Rieke, G.H., et al. 2004, \apjs, 154, 25
\bibitem[Rieke \& Lebofsky(1985)]{rl85} Rieke, G.H. \& Lebofsky, M.J. 1985, \apj,288, 618
\bibitem[Rosenthal, Bertoldi \& Drapatz(2000)]{ros00} Rosenthal, D., Bertoldi, F. \& Drapatz, S. 2000, \aap, 356, 705
\bibitem[Schnee et al.(2005)]{sch05} Schnee, S.L., Ridge, N.A., Goodman, A.A. \& Li, J.G. 2005, \apj, 634, 442
\bibitem[Stepnik et al.(2003)]{ste03} Stepnik, B., et al. 2003, \aap, 398, 551
\bibitem[Storey \& Hummer(1995)]{st95} Storey, J. \& Hummer, D. 1995, \mnras 272, 21
\bibitem[Teixeira \& Emerson(1999)]{tx99} Teixeira, T. C. \& Emerson, J.P. 1999, \aap 351, 292
\bibitem[Volk \& Kwok(1988)]{vk88} Volk, K. \& Kwok, S. 1988,\apj, 331, 435
\bibitem[Weingartner \& Draine(2001)]{wei01} Weingartner, J.C. \& Draine, B.T. 2001, \apj, 548, 296
\bibitem[Wilson et al.(2005)]{wi05} Wilson, B.A., Dame, T.M., Masheder, M.R.W. \& Thaddeus, P. 2005, \aap, 430, 523
\bibitem[Winston et al.(2007)]{wi07} Winston, E., et al. 2007 in prep
\bibitem[Whittet et al.(1993)]{whi93} Whittet, D.C.B., Martin, P.G., Fitzpatrix, E.L., Massa, D. 1993, \apj, 408, 573
\bibitem[Whittet et al.(1997)]{wh97} Whittet, D.C.B., et al. 1997, \apj, 490, 729
\bibitem[Whittet et al.(2001)]{wh01} Whittet, D.C.B., Gerakines, P.A., Hough, J.H. \& Shenoy, S.S. 2001, \apj, 547, 872
\bibitem[Whitney et al.(2003)]{whi03} Whitney, B., Wood, K., Bjorkman, J.E. \& Cohen, M. 2003, \apj, 598, 1079



\end{thebibliography}
\end{document}